%Paper: hep-th/9312188
%From: lerche@nxth04.cern.ch
%Date: Thu, 23 Dec 93 16:27:58 +0100
%Date (revised): Sun, 26 Dec 93 13:48:31 +0100
%Date (revised): Tue, 25 Jan 94 09:35:45 +0100
%Date (revised): Fri, 30 Sep 94 10:08:48 +0100

%%%%%%%%%%%%%%%%%%%%%%%%%%%%%%%%
%%% this paper has 2 uuencoded compressed
%%% postscript figures appended, and also
%%% provides hypertex links if texed with lanlmac.
%%%%%%%%%%%%%%%%%%%%%%%%%  %%%%%%%%%%%%%%%%%%%%%%%%%%%%%%%
% this set of macros is an extension of harvmac
% (9/91 or later) and its hypertex version, lanlmac.
% for info on hypertex, see
% http://xxx.lanl.gov/hypertex/
%%%%%%%%%%%%%%%%%%% %%%%%%%%%%%%%%%%%%%%%%%%%%%
% determine hypertex mode
\newif\iflanl
\openin 1 lanlmac
\ifeof 1 \lanlfalse \else \lanltrue \fi
\closein 1
\iflanl
    \input lanlmac
\else
    \message{[lanlmac not found - use harvmac instead}
    \input harvmac
\fi
\newif\ifhypertex
\ifx\hyperdef\UnDeFiNeD
    \hypertexfalse
    \message{[HYPERTEX MODE OFF}
    \def\hyperref#1#2#3#4{#4}
    \def\hyperdef#1#2#3#4{#4}
    \def\hypernoname{}
    \def\e@tf@ur#1{}
    \def\hep-th/#1#2#3#4#5#6#7{{\tt hep-th/#1#2#3#4#5#6#7}}
    \def\CERN{\address{CERN, Geneva, Switzerland}}
\else
    \hypertextrue
    \message{[HYPERTEX MODE ON}
%hypertex links to xxx.lanl.gov:
  \def\hep-th/#1#2#3#4#5#6#7{
  {\tt hep-th/#1#2#3#4#5#6#7}}
\def\CERN{\address{
Theory Division, CERN, Geneva, Switzerland}}
\fi
%%%%%%%%%%%%%%%%%%%%%%% %%%%%%%%%%%%%%%%%%%%%%%
\newif\ifdraft

\noblackbox
\catcode`\@=11
\newif\iffrontpage
%%%%%%%%%%%%%%%%%%% %%%%%%%%%%%%%%%%%%%%%%%%%%%%%%%%%%%%%%%%%%%%%%
%%%%% sizes, offsets etc
%%%%%%%%%%%%%%%%%%% %%%%%%%%%%%%%%%%%%%%%%%%%%%%%%%%%%%%%%%%%%%%%%
\ifx\answ\bigans
\def\titleft{\titsm}
\magnification=1200\baselineskip=14pt plus 2pt minus 1pt
%
%%%%% unreduced mode: %%%%
%\voffset=0.35truein\hoffset=0.250truein
\advance\hoffset by-0.075truein
\hsize=6.15truein\vsize=600.truept\hsbody=\hsize\hstitle=\hsize
\else\let\lr=L
\def\titleft{\titla}
\magnification=1000\baselineskip=14pt plus 2pt minus 1pt
%
%%%%% reduced mode: %%%%%%%
\hoffset=-0.75truein\voffset=-.0truein
%?\hoffset=-.25truein\voffset=-.0truein
\vsize=6.5truein
\hstitle=8.truein\hsbody=4.75truein
\fullhsize=10truein\hsize=\hsbody
\fi
\parskip=4pt plus 15pt minus 1pt
%
%%%%%%%%%%%%%%%%%%% %%%%%%%%%%%%%%%%%%%%%%%%%%%%%%%%%%%%%%%%%%%%%%
%%%%% figures
%%%%%%%%%%%%%%%%%%% %%%%%%%%%%%%%%%%%%%%%%%%%%%%%%%%%%%%%%%%%%%%%%
\newif\iffigureexists
\newif\ifepsfloaded
\def\epsfcheck{
\ifdraft% to speed up
\input epsf\epsfloadedtrue
\else
  \openin 1 epsf
  \ifeof 1 \epsfloadedfalse \else \epsfloadedtrue \fi
  \closein 1
  \ifepsfloaded
    \input epsf
  \else
\immediate\write20{NO EPSF FILE --- FIGURES WILL BE IGNORED}
  \fi
\fi
\def\epsfcheck{}}
\def\checkex#1{
\ifdraft
\figureexistsfalse\immediate%
\write20{Draftmode: figure #1 not included}
%\figureexiststrue
\else\relax
    \ifepsfloaded \openin 1 #1
	\ifeof 1
           \figureexistsfalse
  \immediate\write20{FIGURE FILE #1 NOT FOUND}
	\else \figureexiststrue
	\fi \closein 1
    \else \figureexistsfalse
    \fi
\fi}
\def\missbox#1#2{$\vcenter{\hrule
\hbox{\vrule height#1\kern1.truein
\raise.5truein\hbox{#2} \kern1.truein \vrule} \hrule}$}
\def\lfig#1{%  this is to call the figure in the text
\let\labelflag=#1%
\def\numb@rone{#1}%
\ifx\labelflag\UnDeFiNeD%
{\xdef#1{\the\figno}%
\writedef{#1\leftbracket{\the\figno}}%
\global\advance\figno by1%
}\fi{\hyperref{}{figure}{{\numb@rone}}{Fig.{\numb@rone}}}}
\def\figinsert#1#2#3#4{%  this inserts the figure
\epsfcheck\checkex{#4}%
\def\figsize{#3}%
\let\flag=#1\ifx\flag\UnDeFiNeD
{\xdef#1{\the\figno}%
\writedef{#1\leftbracket{\the\figno}}%
\global\advance\figno by1%
}\fi
\goodbreak\midinsert%
\iffigureexists
\centerline{\epsfysize\figsize\epsfbox{#4}}%
\else%
\vskip.05truein
  \ifepsfloaded
  \ifdraft
  \centerline{\missbox\figsize{Draftmode: #4 not included}}%
  \else
  \centerline{\missbox\figsize{#4 not found}}
  \fi
  \else
  \centerline{\missbox\figsize{epsf.tex not found}}
  \fi
\vskip.05truein
\fi%
{\smallskip%
\leftskip 4pc \rightskip 4pc%
\noindent\ninepoint\sl \baselineskip=11pt%
{\bf{\hyperdef\hypernoname{figure}{{#1}}{Fig.{#1}}}:~}#2%
\smallskip}\bigskip\endinsert%
}
%
%%%%%%%%%%%%%%%%%%% %%%%%%%%%%%%%%%%%%%%%%%%%%%%%%%%%%%%%%%%%%%%%%
%%%%%  fonts
%%%%%%%%%%%%%%%%%%% %%%%%%%%%%%%%%%%%%%%%%%%%%%%%%%%%%%%%%%%%%%%%%
%%%%%%%%%%%%%%%%%%% %%%%%%%%%%%%%%%%%%%%%%%%%%%%%%%%%%%%%%%%%%%%%%
\font\bigit=cmti10 scaled \magstep1

\font\titla=cmr10 scaled\magstep3
\font\tenmss=cmss10
\font\absmss=cmss10 scaled\magstep1

\newfam\mssfam
\font\footrm=cmr8  \font\footrms=cmr5
\font\footrmss=cmr5   \font\footi=cmmi8
\font\footis=cmmi5   \font\footiss=cmmi5
\font\footsy=cmsy8   \font\footsys=cmsy5
\font\footsyss=cmsy5   \font\footbf=cmbx8
\font\footmss=cmss8
\def\footfont{\def\rm{\fam0\footrm}
\textfont0=\footrm \scriptfont0=\footrms
\scriptscriptfont0=\footrmss
\textfont1=\footi \scriptfont1=\footis
\scriptscriptfont1=\footiss
\textfont2=\footsy \scriptfont2=\footsys
\scriptscriptfont2=\footsyss
\textfont\itfam=\footi \def\it{\fam\itfam\footi}
\textfont\mssfam=\footmss \def\mss{\fam\mssfam\footmss}
\textfont\bffam=\footbf \def\bf{\fam\bffam\footbf} \rm}
\def\tenpoint{\def\rm{\fam0\tenrm}
\textfont0=\tenrm \scriptfont0=\sevenrm
\scriptscriptfont0=\fiverm
\textfont1=\teni  \scriptfont1=\seveni
\scriptscriptfont1=\fivei
\textfont2=\tensy \scriptfont2=\sevensy
\scriptscriptfont2=\fivesy
\textfont\itfam=\tenit \def\it{\fam\itfam\tenit}
\textfont\mssfam=\tenmss \def\mss{\fam\mssfam\tenmss}
\textfont\bffam=\tenbf \def\bf{\fam\bffam\tenbf} \rm}
\ifx\answ\bigans\def\abstractfont{\tenpoint}\else
\def\abstractfont{\def\rm{\fam0\absrm}
\textfont0=\absrm \scriptfont0=\absrms
\scriptscriptfont0=\absrmss
\textfont1=\absi \scriptfont1=\absis
\scriptscriptfont1=\absiss
\textfont2=\abssy \scriptfont2=\abssys
\scriptscriptfont2=\abssyss
\textfont\itfam=\bigit \def\it{\fam\itfam\bigit}
\textfont\mssfam=\absmss \def\mss{\fam\mssfam\absmss}
\textfont\bffam=\absbf \def\bf{\fam\bffam\absbf}\rm}\fi
%
%%%%%%%%%%%%%%%%%%%%%%%%%%%%% %%%%%%%%%%%%%%%%%%%%%%%%%%%%%%%%%
%%%%% footnotes   (adapted from PHYZZX, no hypertext yet)
%%%%%%%%%%%%%%%%%%%%%%%%%%%%% %%%%%%%%%%%%%%%%%%%%%%%%%%%%%%%%%
\def\f@@t{\baselineskip10pt\lineskip0pt\lineskiplimit0pt
\bgroup\aftergroup\@foot\let\next}
\setbox\strutbox=\hbox{\vrule height 8.pt depth 3.5pt width\z@}
\def\vfootnote#1{\insert\footins\bgroup
\baselineskip10pt\footfont
\interlinepenalty=\interfootnotelinepenalty
\floatingpenalty=20000
\splittopskip=\ht\strutbox \boxmaxdepth=\dp\strutbox
\leftskip=24pt \rightskip=\z@skip
\parindent=12pt \parfillskip=0pt plus 1fil
\spaceskip=\z@skip \xspaceskip=\z@skip
\Textindent{$#1$}\footstrut\futurelet\next\fo@t}
\def\Textindent#1{\noindent\llap{#1\enspace}\ignorespaces}
\def\foot{\global\advance\ftno by1%
\attach{\hyperref{}{footnote}{\the\ftno}{\footsymbolgen}}%
\vfootnote{\hyperdef\hypernoname{footnote}{\the\ftno}{\footsymbol}}}%
%   this is for custom footnote marks:
\def\footnote#1{\global\advance\ftno by1%
\attach{\hyperref{}{footnote}{\the\ftno}{#1}}%
\vfootnote{\hyperdef\hypernoname{footnote}{\the\ftno}{#1}}}%
\newcount\lastf@@t           \lastf@@t=-1
\newcount\footsymbolcount    \footsymbolcount=0
\global\newcount\ftno \global\ftno=0
\def\footsymbolgen{\relax\footsym
\global\lastf@@t=\pageno\footsymbol}
\def\footsym{\ifnum\footsymbolcount<0
\global\footsymbolcount=0\fi
{\iffrontpage \else \advance\lastf@@t by 1 \fi
\ifnum\lastf@@t<\pageno \global\footsymbolcount=0
\else \global\advance\footsymbolcount by 1 \fi }
\ifcase\footsymbolcount
\fd@f\dagger\or \fd@f\diamond\or \fd@f\ddagger\or
\fd@f\natural\or \fd@f\ast\or \fd@f\bullet\or
\fd@f\star\or \fd@f\nabla\else \fd@f\dagger
\global\footsymbolcount=0 \fi }
\def\fd@f#1{\xdef\footsymbol{#1}}
\def\space@ver#1{\let\@sf=\empty \ifmmode #1\else \ifhmode
\edef\@sf{\spacefactor=\the\spacefactor}
\unskip${}#1$\relax\fi\fi}
\def\attach#1{\space@ver{\strut^{\mkern 2mu #1}}\@sf}
%
%%%%%%%%%%%%%%%%%%% %%%%%%%%%%%%%%%%%%%%%%%%%%%%%%%%%%%%%%%%%%%%%%
%%%%% References
%%%%%%%%%%%%%%%%%%% %%%%%%%%%%%%%%%%%%%%%%%%%%%%%%%%%%%%%%%%%%%%%%
\newif\ifnref
\def\rrr#1#2{\relax\ifnref\nref#1{#2}\else\ref#1{#2}\fi}
\def\ldf#1#2{\begingroup\obeylines
\gdef#1{\rrr{#1}{#2}}\endgroup\unskip}
\def\nrf#1{\nreftrue{#1}\nreffalse}
\def\doubref#1#2{\refs{{#1},{#2}}}
\def\multref#1#2#3{\nrf{#1#2#3}\refs{#1{--}#3}}
\nreffalse
\def\refout{\listrefs}
%
%%%%%%%%%%%%%%%%%%% %%%%%%%%%%%%%%%%%%%%%%%%%%%%%%%%%%%%%%%%%%%%%%
%%%%%%% eq numbering
%%%%%%%%%%%%%%%%%%% %%%%%%%%%%%%%%%%%%%%%%%%%%%%%%%%%%%%%%%%%%%%%%
\def\eqn#1{\xdef #1{(\noexpand\hyperref{}%
{equation}{\secsym\the\meqno}%
{\secsym\the\meqno})}\eqno(\hyperdef\hypernoname{equation}%
{\secsym\the\meqno}{\secsym\the\meqno})\eqlabeL#1%
\writedef{#1\leftbracket#1}\global\advance\meqno by1}
\def\eqnalign#1{\xdef #1{\noexpand\hyperref{}{equation}%
{\secsym\the\meqno}{(\secsym\the\meqno)}}%
\writedef{#1\leftbracket#1}%
\hyperdef\hypernoname{equation}%
{\secsym\the\meqno}{\e@tf@ur#1}\eqlabeL{#1}%
\global\advance\meqno by1}
%old:
\def\eqnalign#1{\xdef #1{(\secsym\the\meqno)}
\writedef{#1\leftbracket#1}%
\global\advance\meqno by1 #1\eqlabeL{#1}}
%
%%%%%%%%%%%%%%%%%%% %%%%%%%%%%%%%%%%%%%%%%%%%%%%%%%%%%%%%%%%%%%%%%
%%%%%%  macros for titlepage, marginnotes, etc
%%%%%%%%%%%%%%%%%%% %%%%%%%%%%%%%%%%%%%%%%%%%%%%%%%%%%%%%%%%%%%%%%
\def\hsect#1{\hyperref{}{section}{#1}{section~#1}}
\def\hsubsect#1{\hyperref{}{subsection}{#1}{section~#1}}
\def\chap#1{\newsec{#1}}
\def\chapter#1{\chap{#1}}
\def\sect#1{\subsec{#1}}
\def\section#1{\sect{#1}}
\def\\{\ifnum\lastpenalty=-10000\relax
\else\hfil\penalty-10000\fi\ignorespaces}
\def\note#1{\leavevmode%
\edef\@@marginsf{\spacefactor=\the\spacefactor\relax}%
\ifdraft\strut\vadjust{%
\hbox to0pt{\hskip\hsize%
\ifx\answ\bigans\hskip.1in\else\hskip .1in\fi%
\vbox to0pt{\vskip-\dp
%\vskip4pt
\strutbox\sevenbf\baselineskip=8pt plus 1pt minus 1pt%
\ifx\answ\bigans\hsize=.7in\else\hsize=.35in\fi%
\tolerance=5000 \hbadness=5000%
\leftskip=0pt \rightskip=0pt \everypar={}%
\raggedright\parskip=0pt \parindent=0pt%
\vskip-\ht\strutbox\noindent\strut#1\par%
\vss}\hss}}\fi\@@marginsf\kern-.01cm}
\def\titlepage{%
\frontpagetrue\nopagenumbers\abstractfont%
\hsize=\hstitle\rightline{\vbox{\baselineskip=10pt%
{\abstractfont\pubnum}}}\pageno=0}
\frontpagefalse
\def\pubnum{}
\def\pdate{\number\month/\number\yearltd}
\def\makefootline{\iffrontpage\vskip .27truein
\line{\the\footline}
%\vskip -.1truein\line{\pdate\hfil}
\vskip -.1truein\leftline{\vbox{\baselineskip=10pt%
{\abstractfont\pdate}}}
\else\vskip.5cm\line{\hss \tenrm $-$ \folio\ $-$ \hss}\fi}
\def\title#1{\vskip .7truecm\titlestyle{\titleft #1}}
\def\titlestyle#1{\par\begingroup \interlinepenalty=9999
\leftskip=0.02\hsize plus 0.23\hsize minus 0.02\hsize
\rightskip=\leftskip \parfillskip=0pt
\hyphenpenalty=9000 \exhyphenpenalty=9000
\tolerance=9999 \pretolerance=9000
\spaceskip=0.333em \xspaceskip=0.5em
\noindent #1\par\endgroup }
\def\autskip{\ifx\answ\bigans\vskip.5truecm\else\vskip.1cm\fi}
\def\author#1{\vskip .7in \centerline{#1}}

\def\address#1{\ifx\answ\bigans\vskip.2truecm
\else\vskip.1cm\fi{\it \centerline{#1}}}
\def\abstract#1{
\vskip .5in\vfil\centerline
{\bf Abstract}\penalty1000
{{\smallskip\ifx\answ\bigans\leftskip 2pc \rightskip 2pc
\else\leftskip 5pc \rightskip 5pc\fi
\noindent\abstractfont \baselineskip=12pt
{#1} \smallskip}}
\penalty-1000}
\def\endpage{\tenpoint\supereject\global\hsize=\hsbody%
\frontpagefalse\footline={\hss\tenrm\folio\hss}}
\def\ack{\goodbreak\vskip2.cm\centerline{{\bf Acknowledgements}}}
%%%%%%%%%%%%%%%%%%%%%%%%%%%%% %%%%%%%%%%%%%%%%%%%%%%%%%%%%%%%%%
\def\bfone{\relax{\rm 1\kern-.35em 1}}
\def\inbar{\vrule height1.5ex width.4pt depth0pt}
\def\IC{\relax\,\hbox{$\inbar\kern-.3em{\mss C}$}}
\def\ID{\relax{\rm I\kern-.18em D}}
\def\IF{\relax{\rm I\kern-.18em F}}
\def\IH{\relax{\rm I\kern-.18em H}}
\def\II{\relax{\rm I\kern-.17em I}}
\def\IN{\relax{\rm I\kern-.18em N}}
\def\IP{\relax{\rm I\kern-.18em P}}
\def\IQ{\relax\,\hbox{$\inbar\kern-.3em{\rm Q}$}}
\def\IR{\relax{\rm I\kern-.18em R}}
\font\cmss=cmss10 \font\cmsss=cmss10 at 7pt
\def\ZZ{\relax\ifmmode\mathchoice
{\hbox{\cmss Z\kern-.4em Z}}{\hbox{\cmss Z\kern-.4em Z}}
{\lower.9pt\hbox{\cmsss Z\kern-.4em Z}}
{\lower1.2pt\hbox{\cmsss Z\kern-.4em Z}}\else{\cmss Z\kern-.4em
Z}\fi}
\def\a{\alpha}  \def\d{\delta}

\def\L{\Lambda}

\def\cH{{\cal H}} 
 
\def\cL{{\cal L}} \def\cM{{\cal M}}
 \def\cO{{\cal O}}
 
\def\cR{{\cal R}} 
\def\nup#1({Nucl.\ Phys.\ $\us {B#1}$\ (}
\def\plt#1({Phys.\ Lett.\ $\us  {#1}$\ (}
\def\cmp#1({Comm.\ Math.\ Phys.\ $\us  {#1}$\ (}
\def\prp#1({Phys.\ Rep.\ $\us  {#1}$\ (}
\def\prl#1({Phys.\ Rev.\ Lett.\ $\us  {#1}$\ (}
\def\prv#1({Phys.\ Rev.\ $\us  {#1}$\ (}
\def\mpl#1({Mod.\ Phys.\ Let.\ $\us  {A#1}$\ (}
\def\ijmp#1({Int.\ J.\ Mod.\ Phys.\ $\us{A#1}$\ (}
\def\tit#1|{{\it #1},\ }
%
%%%%%%%%%%%%%%%%%%%%%%%%%%%%%%%% %%%%%%%%%%%%%%%%%%%%%%%%%%%%%%
%%%%% misc %%%%
%%%%%%%%%%%%%%%%%%%%%%%%%%%%%%%% %%%%%%%%%%%%%%%%%%%%%%%%%%%%%%

%

\def\ni{\noindent}

\def\bar{\overline}
\def\us#1{\underline{#1}}

\def\hat{\widehat}
\def\hyp{\vrule height 2.3pt width 2.5pt depth -1.5pt}

\def\Coe#1.#2.{{#1\over #2}}
\def\coeff#1#2{\relax{\textstyle {#1 \over #2}}\displaystyle}
\def\coe#1.#2.{\relax{\textstyle {#1 \over #2}}\displaystyle}

\def\shalf{\relax{\textstyle {1 \over 2}}\displaystyle}

\def\to{\rightarrow}
\def\notin{\hbox{{$\in$}\kern-.51em\hbox{/}}}
\def\shdot{\!\cdot\!}

\def\attac#1{\Bigl\vert
{\phantom{X}\atop{{\rm\scriptstyle #1}}\phantom{X}}}

\def\del{\partial}

\def\nex#1{$N\!=\!#1$}

 \def\ie{{\it i.e.}}
\catcode`\@=12
%%%%%%%%% end macros  %%%%%%% %%%%%%%%%%%%%%%%%%%%%%%%%%%%%%
%%%%%%%%%%%%%%%%%%%%%%%%%  %%%%%%%%%%%%%%%%%%%%%%%%%%%%%%%
\def\sc{superconformal\ }

\def\LG{Lan\-dau-Ginz\-burg\ }
\def\nul#1,{{\it #1},}
\def\tvp{\vrule height 3.2pt depth 1pt} %15 3
\def\thp{\vrule height 0.4pt width 0.45em}
\def\ccw#1{\hfill#1\hfill}
\setbox111=\vbox{\offinterlineskip
\cleartabs
\+ \thp&\cr
\+ \tvp\ccw{}&\tvp\cr
\+ \thp&\cr
\+ \tvp\ccw{}&\tvp\cr
\+ \thp&\cr}
%%%%%%%%%%%%%%%%%%%%%%% %%%%%%%%%%%%%%%%%%%%%%%%%%%%
\ldf\mat{D.J.~Gross and A.A.~Migdal, \prl{64} (1990) 717;
M.~Douglas and S.~Shenker, \nup{235} (1990) 635;
E.~Brezin and V.~Kazakov, \plt{236B} (1990) 144.}
\ldf\TOPALG{E.\ Witten, \cmp{117} (1988) 353; \cmp{118} (1988) 411;
\nup340 (1990) 281.}
\ldf\EYtop{T.\ Eguchi and S.\ Yang, \mpl4 (1990) 1693.}
\ldf\MD{M.\ Douglas, \plt238B(1990) 176.}
\ldf\LS{W.\ Lerche and A.\ Sevrin, {\it On the \LG Realization
of Topological Gravities}, preprint CERN-TH.7210/94,
\hep-th/9403183, to appear in Nucl.\ Phys.}
\ldf\topgr{E.\ Witten,  \nup340 (1990) 281; E.\ and H.\ Verlinde,
\nup348 (1991) 457.}
\ldf\TFTmat{R.\ Dijkgraaf and E.\ Witten, \nup342(1990) 486;
R.\ Dijkgraaf and E.\ and H.\ Verlinde, \nup348 (1991) 435;
For a review, see: R.\ Dijkgraaf, \nul{ Intersection theory,
integrable hierarchies and topological field theory}, preprint
IASSNS-HEP-91/91.}
\ldf\Witgr{E.\ Witten, \nup373 (1992) 187. }
\ldf\DVV{R.\ Dijkgraaf, E. Verlinde and H. Verlinde, \nup{352} (1991)
59.}
\ldf\Loss{A. Lossev, \nul{Descendants constructed from matter field
and K. Saito higher residue pairing in Landau-Ginzburg theories
coupled to topological gravity}, preprint TPI-MINN-92-40-T,
\hep-th/9211090.}
\ldf\BGS{B.\ Gato-Rivera and A.M.\ Semikhatov, \plt B293 (1992) 72,
\hep-th/9207004.}
\ldf\LVW{W.\ Lerche, C.\ Vafa and N.P.\ Warner, \nup324 (1989) 427.}
\ldf\cring{D.\ Gepner, \nul{ A comment on the chiral algebras of
quotient superconformal field theories}, preprint PUPT-1130; S.\
Hosono and A.\ Tsuchiya, \cmp136(1991) 451.}
\ldf\EXTRA{B.\ Lian and G.\ Zuckerman, \plt254B (1991) 417; P.\
Bouwknegt, J.\ McCarthy and K.\ Pilch, \cmp145(1992) 541; A.\
Polyakov, \mpl6(1991) 635; S.\ Mukherji, S.\ Mukhi and A.\ Sen, \plt
266B (1991) 337; E.\ Witten, \nup373 (1992) 187; H.\ Kanno and M.\
Sarmadi, preprint IC/92/150; K.\ Itoh and N.\ Ohta, \nup377 (1992)
113; N.\ Chair, V.\ Dobrev and H.\ Kanno, \plt283B (1992) 194; S.\
Govindarajan, T.\ Jayamaran, V.\ John and P.\ Majumdar, \mpl7 (1992)
1063; S.\ Govindarajan, T.\ Jayamaran and V.\ John, \nul{ Chiral
rings and physical states in $c<1$ string theory}, preprint
IMSc-92/30.}
\ldf\Wchiral{K.\ Ito, \plt B259(1991) 73; \nup370(1992) 123; D.\
Nemeschansky and S.\ Yankielowicz, \nul{ N=2 W-algebras,
Kazama-Suzuki models and Drinfeld-Sokolov reduction}, preprint
USC-91-005A;
L.J.\ Romans, \nup369 (1992) 403; W.\ Lerche, D.\ Nemeschansky and
N.P.\ Warner, unpublished.}
\ldf\topw{K.\ Li, \plt B251 (1990) 54, \nup346 (1990) 329; H.\ Lu,
C.N.\ Pope and X.\ Shen, \nup366(1991) 95; S.\ Hosono, \nul{
Algebraic definition of topological W-gravity}, preprint UT-588;
H.\ Kunitomo, Prog.\ Theor.\ Phys.\ 86 (1991) 745.}
\ldf\MS{P.\ Mansfield and B.\ Spence, \nup362(1991) 294.}
\ldf\KS{Y.\ Kazama and H.\ Suzuki, \nup321(1989) 232.}
\ldf\Keke{K.\ Li, \nup354(1991) 711; \nup354(1991)725.}
\ldf\Vafa{C.\ Vafa, \mpl6 (1991) 337.}
\ldf\dis{J.\ Distler, \nup342(1990) 523.}
\ldf\noncritW{A.\ Bilal and J.\ Gervais, \nup326(1989) 222; P.\
Mansfield and B.\ Spence, \nup362(1991) 294; M.\ Bershadsky, W.\
Lerche, D.\ Nemeschansky and N.P.\ Warner, \plt B292 (1992) 35.}
\ldf\BLNWB{M.\ Bershadsky, W.\ Lerche, D.\ Nemeschansky and N.P.\
Warner, \nup401 (1993) 304.}
\ldf\EYQ{T.\ Eguchi, H.\ Kanno, Y.\ Yamada and S.-K.\ Yang, \plt B305
(1993) 235, \hep-th/9302048.}
\ldf\ttstar{S.\ Cecotti and C.\ Vafa, \nup367 (1991) 359.}
\ldf\gep{D.\ Gepner, \nul{Foundations of rational quantum field
theory 1}, preprint CALT-68-1825.}
\ldf\disp{I.\ Krichever, \cmp143 (1992) 415; B.\ Dubrovin, \nup379
(1992) 627, \cmp 145 (1992) 195, \cmp 152 (1993) 539.}
\ldf\Wrev{For reviews on $W$-algebras, see: P.\ Bouwknegt and K.\
Schoutens, Phys.\ Rep.\ 223 (1993) 183; J. \ de Boer, {\it Extended
conformal symmetry in non-critical string theory}, Ph.D.\ thesis,
1993; J.\ Goeree, {\it Higher spin extensions of two-dimensional
gravity}, P.D.\ thesis, 1993; T.\ Tjin, {\it Finite and infinite
$W$-algebras}, P.D.\ thesis, 1993.}
\ldf\genDS{M.\ de Groot, T.\ Hollowood and J.\ Miramontes, \cmp145
(1992) 57; N.\ Burroughs, M.\ De Groot, T.\ Hollowood and J.\
Miramontes, \plt B277 (1992) 89; T.\ Hollowood, J.\ Miramontes and
J.\ Guillen, \nul{Generalized integrability and two-dimensional
gravitation}, preprint CERN-TH-6678-92.}
\ldf\JL{J.\ Lacki, Int.\ J.\ Mod.\ Phys.\ A7 (1992) 4871.}
\ldf\BLX{L.\ Bonora, Q.\ Liu and C.\ Xiong, {\it The integrable
hierarchy constructed from a pair of higher KdV hierarchies and its
associated W algebra}, preprint BONN-TH-94-17.}
\ldf\FS{J.\ Fuchs and C.\ Schweigert,
{\it Level-rank duality of WZW theories and isomorphisms
 of N=2 coset models}, preprint NIKHEF-H-93-16.}
\ldf\KosA{B.\ Kostant, Am.\ J.\ Math.\ 81 (1959) 973.}
\ldf\LW{W.\ Lerche and N.\ Warner, {\it On the Algebraic Structure
of Gravitational Descendants in CP(n--1) Coset Models}, preprint
CERN-TH.7442/94, \hep-th/9409069.}
\ldf\KosB{Theorem by Dale Petersen,
private communication by B.\ Kostant.}
\ldf\flatcoa{K.\ Saito, J.\ Fac.\ Sci.\ Univ.\ Tokyo Sec.\ IA.28
(1982) 775; M.\ Noumi, Tokyo.\ J.\ Math. 7 (1984) 1; B.\ Blok and A.\
Varchenko, \tit Topological conformal field theories and the flat
coordinates| preprint IASSNS-HEP-91/5.} \ldf\flatcob{S.\ Cecotti and
C.\ Vafa, \nup367 (1991) 359; W.\ Lerche, D.\ Smit and N.\ Warner,
\nup372 (1992) 87.}
\ldf\Kons{M.\ Kontsevich, \cmp 147 (1992) 1.}
\ldf\integrMat{S. Kharchev, A. Marshakov, A. Mironov, A. Morozov and
A. Zabrodin, \nup380 (1991) 181; C. Itzkson and J. B. Zuber, Int.\
Journ.\ Mod.\ Phys.\ A7 (1992) 5661; M. Fukuma, H. Kawai and R.
Nakayama, Int.\ Journ.\ Mod.\ Phys.\ A6 (1992) 1385; S. Kharchev, A.
Marshakov, A. Mironov and A. Morozov, \mpl8 (1993) 1047; A.
Marshakov, {\it Integrable structures in matrix models and physics of
2D-gravity}, preprint NORDITA-93/21 P.}
\ldf\superpot{P.\ Fendley, W.\ Lerche, S.\ Mathur and N.P.\ Warner,
\nup348 (1991) 66; W.\ Lerche and N.P.\ Warner, \nup358 (1991) 571.}
\ldf\fusionr{D.\ Gepner, \cmp 141 (1991) 381.}
\ldf\dihedral{D.\ Gepner, \cmp 142 (1991) 433.}
\ldf\BMP{P.\ Bouwknegt, J.\ McCarthy and K.\ Pilch, \nul{
Semi-infinite
cohomology of $W$-algebras}, preprint USC-93/11 and ADP-23-200/M15.}
\ldf\twistKS{W.\ Lerche, \plt252B (1990) 349; T.\ Eguchi, S.\ Hosono
and S.K.\ Yang, \cmp140(1991) 159; T.\ Eguchi, T.\ Kawai, S.\
Mizoguchi and S.K.\ Yang, Rev.\ Math.\ Phys.\ 4 (1992) 329.}
\ldf\JdeB{J.\ de Boer, {\it Extended conformal symmetry in
non-critical string theory}, Ph.D.\ thesis, 1993.}
\ldf\krich{I. Krichever, {\it Topological minimal models and soliton
equations}, Landau Institute preprint.}
\ldf\DS{Drinfel'd and V.~G.~Sokolov,
 Jour.~Sov.~Math. {\bf 30} (1985) 1975.}
\ldf\FMVW{P.\ Fendley, S.\ Mathur C.\ Vafa and N.P.\ Warner, \nup348
(1991) 66.}
\ldf\kac{V.\ Kac and D.\ Petersen, {\it 112 constructions of the
basic representation of the loop group of $E_8$}, in: Symp.\ on
Anomalies, geometry and Topology, eds, W.\ Bardeen and A.\ White,
World Scientific, Singapore 1985.}
\ldf\multiKP{See e.g., V.\ Kac and J.\ van de Leur, {\it
The n-component KP hierarchy and representation theory},
MIT preprint 1993.}
\ldf\qucoho{E.\ Witten, \cmp 118 (1988) 411, \nup 340 (1990) 281; K.\
Intriligator, \mpl6 (1991) 3543; C.\ Vafa, {\it Topological mirrors
and quantum rings}, in: Essays in Mirror Symmetry, ed.\ S.T.\ Yau,
1992; V.\ Sadov, {\it On the equivalence of Floer's and quantum
cohomology}, preprint HUTP-93/A027; E.\ Witten, {\it The Verlinde
algebra and the cohomology of the grassmannian}, preprint
IASSNS-HEP-93/41.}
\ldf\kamig{V.\ Kazakov and A.\ Migdal, \nup397 (1993) 214.}
\ldf\wconstr{M.\ Fukuma, H.\ Kawai and R.\ Nakayama, Int.\ J.\
Mod.\ Phys.\ A6 (1991) 1385; R.\ Dijkgraaf and E.\ and H.\ Verlinde,
\nup 356 (1991) 574; A.\ Marshakov, A.\ Mironov and A.\ Morozov,
\plt 274 (1992) 280.}
\ldf\wbrs{M.\ Bershadsky, W.\ Lerche, D.\ Nemeschansky and N.P.\
Warner, \plt B292 (1992) 35, \hep-th/9207067; E.\
Bergshoeff, A.\ Sevrin and X.\ Shen, \plt B296 (1992) 95,
\hep-th/9209037; J. de Boer and J. Goeree, \nup405 (1993) 669,
\hep-th/9211108.}
\ldf\virnul{M.\ Bauer, P.\ Di Francesco, C.\ Itzykson and J.\ Zuber,
\plt B260(1991) 323, \nup362(1991) 515.}
\ldf\diFKu{P.\ Di Francesco and D.\ Kutasov, \nup342(1990)589.}
\ldf\period{T.\ Eguchi, Y.\ Yamada and S.\ Yang, \mpl8(1993) 1627.}
\ldf\taka{K.\ Takasaki, {\it Integrable hierarchies underlying
topological Landau-Ginzburg models of $D$-type}, preprint
KUCP-0061/93.}
\ldf\OSSPvN{H.\ Ooguri, K.\ Schoutens, A.\ Sevrin and P.\ van
Nieuwenhuizen, \cmp145 (1992) 515.}
\ldf\BFFOW{J.\ Balog, L.\ Feher, P.\ Forgacs, L.\ O'Raifeartaigh
and A.\ Wipf, \plt B244 (1990) 435.}
\ldf\othergrad{L.\ Feher, L.\ O'Raifeartaigh, P.\ Ruelle, I.\ Tsutsui
and A.\ Wipf, Ann.\ Phys.\ 213 (1992)~1; L.\ Feher, J.\ Harnad and
I.\ Marshall, \cmp154 (1993) 181.
}
%%%%%%%%%%%%%%%%%%%%%%%%%%%% %%%%%%%%%%%%%%%%%%%%%%%%%%%%%

\def\LG{Lan\-dau-Ginz\-burg\ }
\def\cmq#1#2#3{\cM^{(#1)}_{#2,{#3}}}
\def\ga#1{\sigma_{#1}}
\def\cph#1#2{\widehat{CP}_{#1}^{\lower2pt\hbox{$\scriptstyle(#2)$}}}
\def\rnk#1#2{\cR_x^{(#1,#2)}}
\def\rnkg#1#2{\cR_{x,\sigma}^{(#1,#2)}}
\def\wnk#1#2{W^{(#1,#2)}}
\def\cph#1#2{{\rm CP}_{\!#1,#2}^{{\rm top}}}
\def\rnk#1#2{\cR^x_{#1,#2}}
\def\rnkg#1#2{\cR^{x,\sigma}_{#1,#2}}
\def\wnk#1#2{W_{#1,#2}}
\def\GH{$G/H$}
\def\L{\Lambda}
\def\dx#1{\del_{x_{#1}}}
\def\Lz{\L^{(z)}}
\def\O{\Omega}
\def\bfone{{\bf 1}}
\def\wth{\widetilde\cH}
%
%\draft
%
\def\pubnum{
\hbox{CERN-TH.6988/93}
\hbox{hep-th@xxx/9312188}
\hbox{Revised}}
\def\pdate{}
\titlepage
\vskip 1.cm
\title
{Generalized Drinfeld-Sokolov Hierarchies,\break Quantum Rings, and
W-Gravity}
\vskip .3cm
\author{\ W.$\,$Lerche}
\CERN
\vskip0. cm
\abstract{
We investigate the algebraic structure of integrable hierarchies
that, we propose, underlie models of $W$-gravity coupled to matter.
More precisely, we concentrate on the dispersionless limit of the
topological subclass of such theories, by making use of a
correspondence between Drinfeld-Sokolov systems, principal $s\ell(2)$
embeddings and certain chiral rings. We find that the integrable
hierarchies can be viewed as generalizations of the usual matrix
Drinfeld-Sokolov systems to higher fundamental representations of
$s\ell(n)$. Accordingly, there are additional commuting flows as
compared to the usual generalized KdV hierarchy. These are
associated with the enveloping algebra and account
for degeneracies of physical operators. The underlying Heisenberg
algebras are nothing but specifically perturbed chiral rings of
certain Kazama-Suzuki models, and have an intimate connection with
the quantum cohomology of grassmannians. Correspondingly, the Lax
operators are directly given in terms of multi-field superpotentials
of the associated topological LG theories. We view our construction
as a prototype for a multi-variable system and suspect that it might
be useful also for a class of related problems.
}

\vfil
%\vskip 1.cm
\ni CERN-TH.6988/93\hfill\break
\ni December 1993
\endpage
\baselineskip=14pt plus 2pt minus 1pt
%%%%%%%%%%%%%%%%%%%%%%%%%%%%% %%%%%%%%%%%%%%%%%%%%%%%%%%%%%%%

\chapter{Introduction}
%%%%%%%%%%%%%%%%%%%%%%%%%%%%% %%%%%%%%%%%%%%%%%%%%%%%%%%%%%%%

There is considerable interest in the study of $2d$ conformal matter
coupled to gravity. In particular, minimal models of type $(p,q)$
coupled to gravity can be described in terms of matrix models \mat,
whose dynamics are governed by the (generalized) KdV hierarchy
\multref\MD\TFTmat\integrMat. Recently, an infinite series of new
theories based on $W$-matter coupled to $W$-gravity has been
constructed \doubref\wbrs\BLNWB. Since these theories appear to be on
a footing similar to ordinary gravity, one might suspect that there
exists an infinite sequence of new types of matrix models too,
governed by certain integrable hierarchies.

The present paper is a first attempt to identify the algebraic
structure
of these integrable systems.\foot{To prevent confusion at an early
stage, note that we will not be talking about the well-known
classical, Poisson bracket $W$-algebra of the generalized KdV
hierarchy, which arises already in ordinary gravity. What we are
about to discuss is an extension that sits on top of this structure.
See \hsect4\ for an expos\'e of the general scheme.} We certainly
would not be surprised if these ``new'' hierarchies ultimately turn
out to be equivalent to some already known (say, multi-component
\multiKP) generalization of the KP-hierarchy. It seems that in
particular the work in refs.\ \multref\JL{\genDS\JdeB}\BLX\ is, in
one way or another, related to ours. However, so far no connection of
these integrable systems to concrete models of $2d$-gravity was made,
and such a connection will be one of the virtues of our construction.

Since we do not know the precise structure of the suspected new
matrix models as yet, we cannot derive the hierarchies from first
principles, \ie, in the way the KdV hierarchy can be extracted from
the usual matrix models \refs{\MD{--}\integrMat}. Rather, we will
follow an indirect route by making use of a connection to topological
\LG theory \Vafa. For this, we will need to focus on a sub-class of
theories, that is, on the topological models of type $(1,q)$, in the
dispersionless limit (we also restrict to the ``small phase space''
of primary matter couplings, where this limit is exact). We reserve
the extension to general $(p,q)$  for future work.

More specifically, recall that the ordinary $(1,q)$ Virasoro minimal
models coupled to gravity are closely related
\refs{\Keke{,}\DVV{,}\BGS{,}\BLNWB}\ to the twisted
\doubref\TOPALG\EYtop\ \nex2 superconformal minimal models of type
$A_{k+1}$ (where $k\!=\!q\!-\!2$). The point is \DVV\ that the \LG
potentials of these topological models, $W(x,g)$, are very similar to
the Lax operators $L(D,g)$ of the $(1,q)$ type of KdV hierarchy. In
fact, in the dispersionless limit \disp, where $D\to x$, $L$ and $W$
coincide and the KdV flow equations, which determine the dynamics of
the $(1,q)$ matrix models, become equivalent to the Gau\ss-Manin
equations \doubref\flatcoa\flatcob, which determine the correlation
functions of the LG models \DVV.

Our strategy will be to first make an inspired guess about the
structure of the integrable systems related to $W$-gravity, guided by
the above-mentioned relationship between topological LG theory and
KdV integrable systems. In particular, at the heart of our
construction will be the generalization of the relationship between
\LG superpotential and dispersionless Lax operator to several
variables.

We then would need to verify that the associated flow equations have
the correct solutions. While we were not able to prove this in
generality, we checked explicitly for a couple of examples related to
$W_3$ that the flow equations indeed reproduce at least the correct
flat coordinates of the topological matter models. From the logic
of the scheme it is fairly obvious that this feature should work
in general. Note that this is, ultimately, more than just
determining flat coordinates in topological matter systems: the
integrable systems are supposed to describe not only the
matter sector, but also the gravitational sector of a theory.

More specifically, after having recalled the relationship between
$W$-matter-gravity systems and certain \nex2 superconformal coset
models, we will present in \hsect2\ a useful characterization of the
chiral rings of these coset models. It turns out that principal
embeddings of $s\ell(2)$ play a crucial r\^ole here, quite analogous
to the well-known r\^ole they play in Drinfeld-Sokolov systems. In
\hsect3, we will then first reformulate the relationship between the
KdV hierarchy and the ordinary \nex2 minimal models in a way that is
most suitable for later generalization. We will in particular note a
close correspondence between the KdV matrix Lax operators and the
minimal model ring structure constants, and subsequently employ this
correspondence for the generalization to $W_3$. Section 3 will be
concluded with an explicit example and with a brief discussion about
some observations we made when we computed various examples. In
particular, we will find that the generalized flow equations are just
the integrability conditions for the existence of a ``prepotential''.
Finally, in \hsect4\ we will review the philosophy of our
construction and make as well some more speculative remarks.

\goodbreak
%%%%%%%%%%%%%%%%%%%%%%% %%%%%%%%%%%%%%%%%%%%%%%%%%%%%%%%%%%%%
\chapter{Topological W$_{\!{\textstyle{\bf n}}}$ models}
\section{Equivariant cohomology of Kazama-Suzuki models and
W-gravity}
%%%%%%%%%%%%%%%%%%%%%%% %%%%%%%%%%%%%%%%%%%%%%%%%%%%%%%%%%%%%

\ni The physical theories in focus are tensor products \noncritW
$$
\cmq npq\ \equiv\
\Big[\,M^{(n)}_{p,q}\otimes W_n\hyp{\rm Liouville}
\otimes{\rm ghosts}\,\Big]\ \ \ {\rm with}\ p=1\ ,
\eqn\defmod
$$
where $M^{(n)}_{p,q}$ denotes a $(p,q)$ type minimal model of the
$W_n$ algebra \Wrev, and ``$W_n\hyp{\rm Liouville}$'' denotes an
$s\ell(n)$ Toda theory that describes the coupling to $W_n$-gravity
(ordinary gravity corresponds to $n\!=\!2$). There is strong evidence
\BLNWB\ that the sub-class of $(1,q)$ type of theories is equivalent
to topological matter coupled to topological $W$-gravity \topw:
$$
\cmq n1{q=k+n}\ \cong\ \Big[\,\cph{n-1}k\otimes {\rm
topological\ } W_n\hyp{\rm gravity}\,\Big]
\eqn\deftop
$$
Here, $\cph{n-1}k$ denotes a minimal topological $W_n$ matter model
\Wchiral\ at level $k$, which is nothing but the twisted version
\twistKS\ of a \nex2 \sc\ coset model \KS\ based upon ${SU(n)_k\over
U(n-1)}$, with anomaly $c={3k(n-1)\over n+k}$. The physical spectrum
of \deftop\ is given by a chiral ground ring \Witgr, which consists
of two parts. The matter part is given by the primary chiral ring of
the matter model $\cph{n-1}k$, which is generated by fields $x_i$,
$i=1,\dots,n-1$ (with $U(1)$ charges $q(x_i)=\coeff i{n+k}$) and
which can be represented by
$$
\rnk nk\ =\ \Big\{\prod_{i=1}^{n-1}(x_i)^{m_i}\ ,\sum m_i\leq
k\,\Big\}\ .
\eqn\xring
$$
The remaining part consists of the gravitational descendants and is
generated by operators $\ga i$, $i=1,\dots,n-1$ (with $U(1)$ charges
$q(\ga i)=1$). The complete\foot{Originally \BLNWB, only a subset of
these ring elements was considered, namely the subset that is
associated with generators with charges $q(\ga i)=i$.} ground ring
thus has the form
$$
\rnkg nk\ =\ \rnk nk \otimes \Big\{\prod_{i=1}^{n-1}(\ga i)^{l_i}\ ,\
l_i=0,1,2,\dots\,\Big\}\ .
\eqn\fullring
$$

It is well-known that the matter models $\cph{n-1}k$ have a \LG
realization, where the LG fields represent the ground ring
generators $x_i$ above. The superpotentials $\wnk nk(x_i)$ were
explicitly given in \doubref\LVW\fusionr\ and can be compactly
characterized by the following generating function:
$$
-\log\Big[\sum_{i=1}^{n-1}(-t)^ix_i\,\Big]\ =\
\sum_{k=-n+1}^{\infty}t^{n+k}\,\wnk{n}k(x)\
\eqn\genallW
$$
These superpotentials obey the following recursive identity
that will be important in the following:
$$
\eqalign{
\wnk{n}k(x)\ &=\
{1\over n+k}\,\nabla_x\,\wnk{n}{k+1}(x)\ ,\cr
\nabla_x \ &\equiv \ \sum_{i=1}^{n-1}(n-i)\,x_{i-1}{\del\over\del
{x_i}}\cr
}\eqn\relallW
$$
This identity can easily be proven by
observing that the differential operators $\nabla_x$ and
$$
\nabla_t \ = \ t^2\,{\del\over\del t} + (n-1)\,t
$$
give the same result when acting on the LHS of \genallW.

It is also known \BLNWB\ that by changing the cohomological
definition
of topological matter models $\cph{n-1}k$, i.e., by requiring
equivariant cohomology \topgr, the gravitational sector can be
represented
entirely in terms of the matter sector:
$$
\Big[\,\cph{n-1}k\otimes {\rm
topological\ } W_n\hyp{\rm gravity}\,\Big]\ \cong
\cph{n-1}k\attac{{{\rm equivariant}\atop{\rm cohomology}}}\ .
\eqn\deftop
$$
The gravitational sector is thus implicitly contained in the
topological matter model, and for the LG representation this means
that (in equivariant cohomology) one can represent the gravitational
descendants in terms of LG fields as well. In particular, for
ordinary minimal models $\cph1k\sim A_{k+1}$, one has the following
LG representatives \doubref\Loss\EYQ\
$$
\ga l(x_i)\ \equiv\ (\ga1)^l(x_1)^i\ \cong\ x^{i+l(k+2)}_1\ ,
$$
where $x_1^{(l+1)(k+2)-1}\equiv (\ga1)^l\del_x W_{2,k}$ are
BRST trivial. The equivariant LG spectrum for $k=2$ is depicted in
\lfig\figone.
%%%%%%%%%%%%%
\figinsert\figone{Gravitational chiral ring associated with the model
$\cph12\sim A_3$. The open dots describe null fields, and the
repetitions of the matter subring $\{1,x,x^2\}$ correspond to the
gravitational excitations. Mathematically, the picture represents the
highest weight representations of $\widehat{s\ell}(2)$ at level 2;
the matter subring corresponds to the subset of integrable
representations. It also represents the spectrum of type $(1,k+2)$
KdV flows.}{.7in}{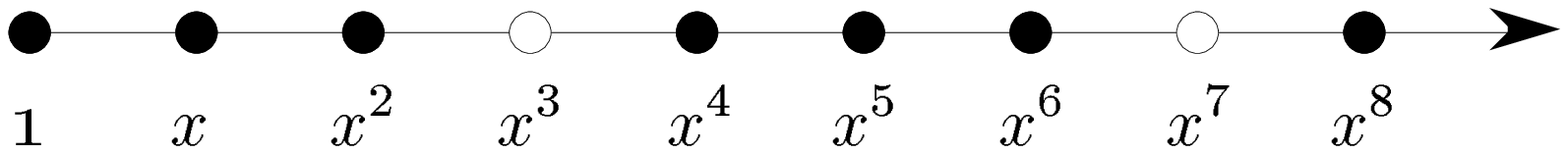}
%%%%%%%%%%%%%
The situation is analogous for the models $\cph{2}k$ associated with
$W_3$-gravity, on which we will primarily focus in the following. It
is
convenient choose a specific ring basis in these models, namely the
eigenbasis corresponding to the underlying $W_3$-symmetry \LS. This
basis is given by the LG polynomials
$$
\Phi^{m_1,m_2}(x_1,x_2)\ =\ {x_2}^{m_2}\Big({\del\over\del
x_1}\wnk3{m_1-2}(x_1,x_2)\Big)\ ,\ \  m_{1,2}\geq0\ ,
\eqn\basis
$$
so that the matter chiral ring \xring\ can be represented as
$$
\rnk 3k\ =\ \Big\{\,\Phi^{m_1,m_2}(x_1,x_2)\,,\ m_1+m_2\leq
k\,\Big\}\ .
\eqn\xringLG
$$
By writing $\lambda^{m_1,m_2}=m_1\lambda_1+m_2\lambda_2$ (where
$\lambda_{1,2}$ are the fundamental weights of $s\ell(3)$), the ring
elements \xringLG\ can be associated with the integrable highest
weights of $\widehat{s\ell}(3)$ at level $k$. If one requires
equivariant cohomology (in order to incorporate the $W$-gravitational
descendants), one finds \doubref\LS\LW\ that the LG polynomials
$\Phi^{m_1,m_2}$ corresponding to {\it all} (in general not
integrable) highest weights become physical. Recalling the structure
of the affine highest weights, one can thus represent the complete
spectrum in terms of LG variables as follows:
$$
\rnkg 3k\ =\ \Big\{\,\Phi^{m_1,m_2}(x_1,x_2)\,:\ \,
(\lambda^{m_1,m_2}+\rho)\cdot \a_i\, \not=\, 0\, {\rm mod}\,
(k\!+\!3)\,\Big\}\
\eqn\fullringLG
$$
(here, $\rho$ and $\a_i$ denote Weyl vector and
roots of ${s\ell}(3)$). We depicted the spectrum for $k=2$
in \lfig\figtwo.
%%%%%%%%%%%%%
\figinsert\figtwo{Landau-Ginzburg representation
$\Phi^{m_1,m_2}(x_1,x_2)$ of the gravitational chiral ring
associated with the model $\cph22$. The open dots describe null
fields, and the repetitions of the matter ``triangle'' correspond to
the $W_3$-gravitational excitations. The picture represents the
highest weight representations of $\widehat{s\ell}(3)$ at level
2.}{2.2in}{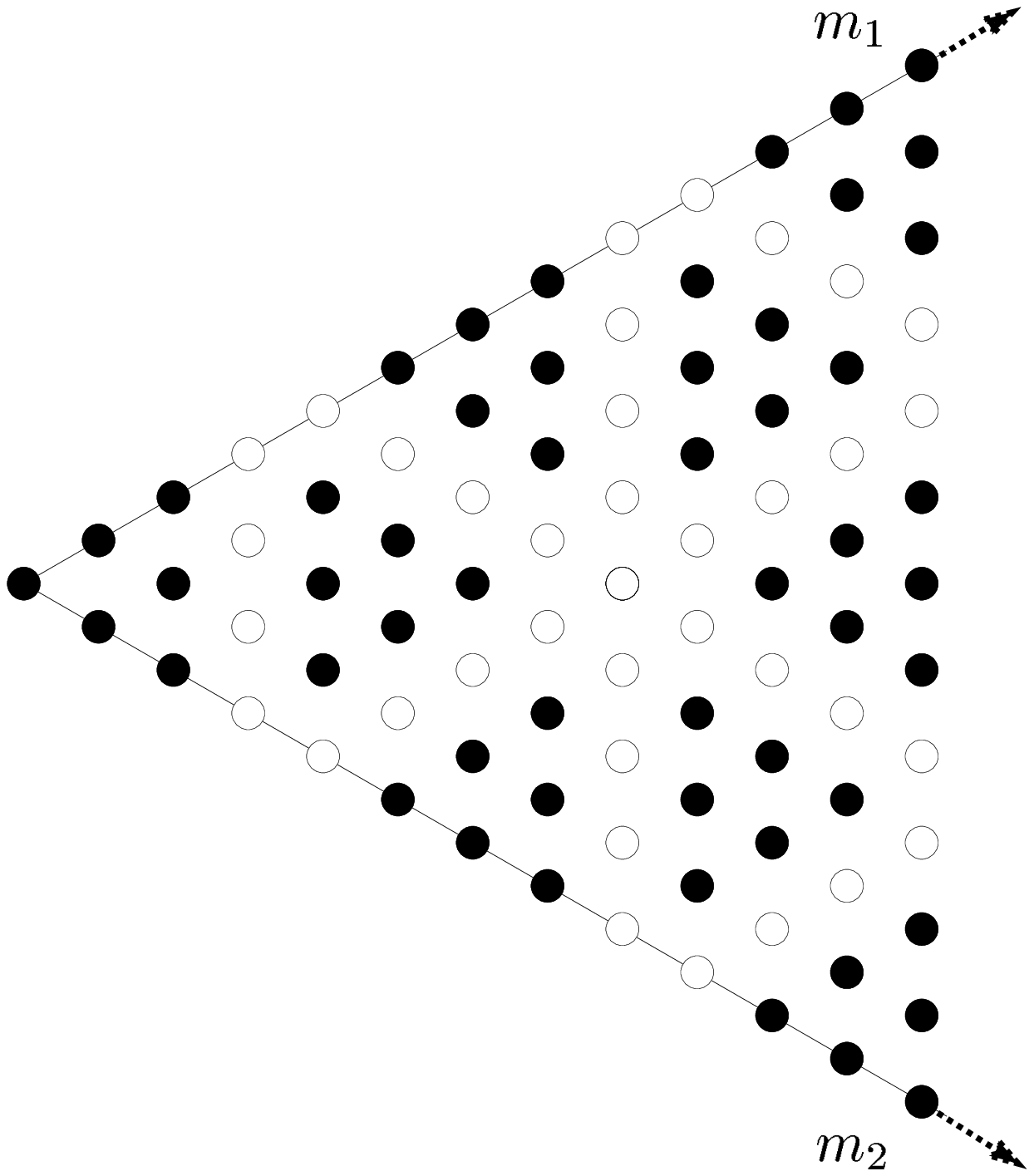}
%%%%%%%%%%%%%

It is clear that \lfig\figone\ represents simultaneously the spectrum
of type $(1,k+2)$ KdV flows, which describe the dynamics of the
ordinary matter-gravity systems $\cph{1}k$. Each black dot
corresponds to a flow hamiltonian, and each open dot to a trivial
flow. In the following, we attempt to find integrable systems that
pertain in an analogous way to spectra of $W$-gravity systems, like
to the spectrum in \lfig\figtwo.

\goodbreak
%%%%%%%%%%%%%%%%%%%%%%% %%%%%%%%%%%%%%%%%%%%%%%%%%%%%%%%%%%%%
\section{Chiral rings and principal embeddings of SU(2)}
%%%%%%%%%%%%%%%%%%%%%%% %%%%%%%%%%%%%%%%%%%%%%%%%%%%%%%%%%%%%

We like to characterize in this section the algebraic structure of
the chiral rings $\rnk nk$, which will turn out to be closely related
to the structure of the generalized KdV flow hamiltonians.

There is a well-known duality symmetry \refs{\KS{,}\dihedral{,}\FS}\
among some of the \nex2 superconformal coset models, and in
particular there is an isomorphism between the coset models based on
$SU(m+l)_k\over SU(m)\times SU(n)\times U(1)$ for all permutations of
$m,l,k$. This allows to associate the \nex2 $W_n$ minimal models
$\cph{n-1}k$ (based on $G_k/H={SU(n)_k\over U(n-1)}$) also with the
grassmannians $G_1/H\!=\!{SU(n+k-1)_1\over SU(n-1)\times SU(k)\times
U(1)}$. The point is that for any such model where $G/H$ is a
hermitian symmetric space, $G$ is simply laced and the level of $G$
is equal to one, there exist powerful theorems about the structure of
the chiral rings.

Specifically, it was noticed \doubref\LVW\cring\ that in such models,
the chiral, primary fields are in one-to-one correspondence with the
Lie-algebra cohomology, and the primary chiral ring of a given such
model based on $G/H$ is isomorphic to the Dolbeault cohomology ring
of \GH. In particular,
$$
\rnk nk \ \cong\ H^*_{\bar\del}\big(\,\coeff{SU(n+k-1)}{SU(n-1)\times
SU(k)\times U(1)}\,,\IR\,\big)\ .
\eqn\ringiso
$$
That is, a chiral, primary field with \nex2 $U(1)$ charge $p\over
g+1$ can be thought of as an element of $H^{p,p}_{\bar\del}(G/H,\IR)$
($g$ denotes the Coxeter number of $G$; for $G\!=\!SU(m)$,
$g\!=\!m$). A characterization of these chiral rings in terms of
Chern classes of vector bundles over \GH\ was given in
\doubref\superpot\fusionr.

In addition, it was shown that the chiral ring of any such
theory has a close relationship to a particular fundamental
representation, $\Xi$, of $G$. The primary, chiral fields are
one-to-one to the weights of $\Xi$ and their $U(1)$ charges are
given, up to a uniform shift, by the dot product of the corresponding
weight with the Weyl vector $\rho_G\equiv{1\over2}\sum_{\left({\rm
{positive\atop roots}}\right)}\a$. The highest weight of $\Xi$ (also
denoted by $\Xi$) is defined by the fundamental weight of $G$
corresponding to the node of the $G$-Dynkin diagram that defines the
embedding of the $U(1)$ factor in $G$. The possible choices of $U(1)$
factors correspond to the Dynkin nodes which have Kac weight equal to
one, i.e., the allowed representations $\Xi$ are the level one
representations of affine-$G$. For the specific models under
consideration, $\cph{n-1}k$, $\Xi$ is given by the $(n-1)$-th
fundamental representation of $G=SU(n+k-1)$, with highest weight
$\Xi=(0,\dots,0,1,0,\dots,0)$\ (where $"1"$ appears at the $(n-1)$-th
entry).

This leads to a further characterization of the chiral rings that is
most useful in the present context. Crucial to this is the principal
$SU(2)$ subgroup $S\subset G$ generated by
$$
\eqalign{
I_+\ &=\
\sum_{{\rm{simple\atop roots\ \alpha}} }a_\alpha^{(1)}\, E_\alpha\cr
I_-\ &=\
\sum_{{\rm{simple\atop roots\ \a}}}a_\alpha^{(-1)} \,E_{-\alpha}\cr
I_0\ &=\ \rho_G\cdot H\ ,}
\eqn\Sdef
$$
where $E_{\pm\a}$ are, as usual, the generators of $G$ in the
Cartan-Weyl basis and $a_\alpha^{(\pm1)}$ are coefficients such that
$[I_+,I_-]= I_0$. One can always take $a_\alpha^{(1)}\equiv 1$, and
this is what we will assume henceforth. The particular choice of
$I_0$ induces the ``principal'' gradation of the generators of $G$:
the $I_0$ charge of a generator $E_\a$ is given by $p=\rho_G\cdot\a$.
It is a well-known mathematical fact \KosA\ that the possible values
of $|p|$ are just given by the exponents $m_i, i=1,\dots,\ell$, of
$G$. One can group the generators into sets of equal $I_0$ grade, and
build the following linear combinations:
$$
\eqalign{
\L_p\ =\ &\sum_{\{\a:\rho_G\cdot\a=p\}}a_\a^{(p)} E_\a\ ,\ \ \
{\rm\  for\ each\ } p\in\{\pm m_1,\pm m_2,\dots,\pm m_\ell\}\ ,\cr
{\rm with}\ \ &\big[\L_p,I_0\,\big]\ =\ p \,\L_p\cr}
\eqn\Fpdef
$$
(where $\L_1=I_+, \L_{-1}=I_-$). The coefficients $a_\a^{(p)}$ (for
$|p|>1$) are determined from \Sdef\ by requiring:
$$
\big[\L_p,\L_q\,\big]\ =\ 0 {\rm\ \ \ for\ \ \ }
\cases{ p>0 , q>0\cr p<0 , q<0 \cr}
\eqn\Fpcomm
$$
For $G=SU(m)$, one can take $\L_p=\sum_{i=1}^{m-p}E_{e_i-e_{i+p}}\
(p=1,2,\dots,m-1)$.

\ni A useful theorem \KosB\ can now be formulated as follows:
\hfill\break {\it In the representation $\Xi$ of $G$ defined above,
the matrices $\L_p$ (with $p>0$)\foot{One can equally well restrict
to $p<0$.} generate an algebra that is isomorphic to the
cohomology ring $H^*_{\bar\del}(G/H,\IR)$}.

It follows that the matrices $\L_p, p>0$ represent
the \LG fields $x_p$ and that they are identical to the (unperturbed)
ring structure constants:
$$
{(C_p)_j}^k(0)\ =\ {(\L_p)_j}^k\ \ \ \ \ \ (i,k=1,\dots,{\rm
dim}\Xi)\ .
\eqn\ringC
$$
Ring multiplication is represented by simple matrix
multiplication. Generic ring elements are given by polynomials in
$\L_p$, which do not necessarily belong to the algebra of $G$; by
definition, they belong to the enveloping algebra of $G$. The $U(1)$
charge of a ring element is equal to its $I_0$ grade in units of
$1/(g+1)$. Obviously, $\L_1\equiv I_+$ is always a generator of the
ring, and this corresponds to the fact that in each coset model,
there is a unique \LG field of lowest $U(1)$ charge,
$q(x_1)=\coeff1{g+1}$.

In a sense, the powers of $\L_1$ always describe the chiral ring of
an ordinary \nex2 minimal model, but if $\Xi$ is larger than the
defining representation of $G$, there exist additional, independent
commuting matrices (proportional to $\L_{p\not=1}$) that represent
additional ring elements. It is precisely this point of view that we
will take later to describe the integrable hierarchies: if one
considers representations higher than the defining representation,
then there will exist extra, commuting flows besides the ordinary KdV
flows.

In general, various $\L_p$ can be expressed in terms of powers of
other ${\L_q}$ so that they are not independent generators; which
$\L_p$ are independent for a given group $G$ depends on the
representation $\Xi$, \ie, on the choice of $H$. We will denote in
the following the independent generators by $\L_i$, $i=1,\dots,M$.
For the grassmannians associated with $W_n$ gravity, there are
$M\equiv(n-1)$ independent generators with degrees $1,\dots,(n-1)$,
in accordance with \xring. The independent generators represent the
independent \LG fields $x_i$, and satisfy certain relations. There is
always a subset of relations that generate all other ones; it is
non-trivial \LVW\ that these generating vanishing relations are
always of the form $\del_{x_i}W(x)=0$ for some $W$ and thus can be
interpreted as the equations of motion of a \LG theory with
superpotential $W$.

The fact that $\L_i$ represents the \LG field $x_i$, when
acting on the space of primary chiral fields, can be expressed by the
following matrix system:
$$
\Big[\,x_i\,{\bf1} - \L_i\,\Big] \cdot \Psi\ =\
0\ , \ \ \ \ \Psi_j\in\cR^x\ .
\eqn\mateq
$$
The components of the solution vector $\Psi$ represent the weights of
$\Xi$ and can be recursively solved for. The solution $\Psi_j(x_i)$
in terms of polynomials in the \LG fields thus gives the precise
relationship between the weights of $\Xi$ and the \LG fields. When
all unknown $\Psi_j$ are eliminated in favor of the first component
$\Psi_0\equiv1$, the system \mateq\ is reduced to a set of vanishing
relations for the \LG fields $x_i$. One can thus regard \mateq\ as
fundamental equations characterizing the chiral ring. As we will see,
$\L_i$ are closely related to Lax operators of integrable systems.

To give some examples, consider first the models $\cph 1k$, which are
the same as the ordinary twisted \nex2 minimal models of type
$A_{k+1}$. These models can be associated to cosets $SU(k+1)_1/U(k)$,
and $\Xi$ is the defining representation of $SU(k+1)$. There is just
one independent generator of the chiral ring, which can be
represented by the familiar, ubiquitous $(k+1)\times(k+1)$
dimensional matrix
$$
\L_1\ \equiv \ I_+\ =\ \pmatrix
{ 0 & 1 & 0 & \dots & 0 \cr
  0 & 0 & 1 & \dots & 0 \cr
  \vdots & \vdots & \vdots & \ddots &  \vdots \cr
0 & 0 & 0 & \dots & 1 \cr
0 & 0 & 0 & \dots & 0 \cr}\ \ ,
\eqn\step
$$
in terms of which the other $\L_p$ are given by
$$
\L_p\ =\  (\L_1)^p\ , \ \ \ \ p=1,2,\dots,k\ .
\eqn\Fpdep
$$
The vanishing relation is
$$
(\L_1)^{k+1}\ =\ 0\ ,
\eqn\Avanrel
$$
and this corresponds to the vanishing relation in $H^*({\rm CP}_k)$,
and to the equation of motion of a LG model with
superpotential $\wnk2k(x,0)=\coeff1{k+2}x^{k+2}$. The matrix
equation $[x{\bf1}-\L_1]\shdot\Psi=0$ has as solution vector
$$
\Psi\ =\ (1,x,x^2,\dots,x^k)^t \ ,\qquad \Psi_j\in\rnk 2k\ ,
\eqn\ansolvec
$$
and leads to
$$
x^{k+1}\Psi_0\ =\ 0\ .
$$
This is just the characteristic equation
$$
0\ =\ {\rm det}[x{\bf1}-\L_1]\ =\ x^{k+1}\ ,
\eqn\chareq
$$
and thus is the same as the relation \Avanrel\ that the matrix
$\L_1$ itself satisfies.

We now turn to the more interesting models $\cph 2k$, on which we
will primarily focus in the paper. These topological minimal $W_3$
models are associated with the grassmannians $G_1/H={SU(k+2)_1\over
SU(k)\times SU(2)\times U(1)}$, where $\Xi$ is the
${\lower2pt\copy111}\,$-representation $(0,1,0,\dots,0)$, with
dim$\Xi=\shalf(k+1)(k+2)={\rm dim}\rnk 3k$. The independent ring
generators are $x_1\cong\L_1$ and $x_2\cong\L_2$, with
$\L_2\not=(\L_1)^2$ (for $k\geq2$).

For example, for $k=3$ one has in the ten dimensional representation
of $SU(5)$:
\def\mi#1{\!\!{\fiverm#1}\!}
\def\hyp{\vrule height 1.8pt width 2.5pt depth -1.5pt}
$$\L_1=
\pmatrix{ \mi{0} & \mi{1} & \mi{0} & \mi{0} & \mi{0} & \mi{0} &
\mi{0} & \mi{0} & \mi{0} & \mi{0} \cr \mi{0} & \mi{0} & \mi{1} &
\mi{1} & \mi{0} & \mi{0} & \mi{0} & \mi{0} & \mi{0} & \mi{0} \cr
\mi{0} & \mi{0} & \mi{0} & \mi{0} & \mi{1} & \mi{0} & \mi{0} & \mi{0}
& \mi{0} & \mi{0} \cr \mi{0} & \mi{0} & \mi{0} & \mi{0} & \mi{1} &
\mi{1} & \mi{0} & \mi{0} & \mi{0} & \mi{0} \cr \mi{0} & \mi{0} &
\mi{0} & \mi{0} & \mi{0} & \mi{0} & \mi{1} & \mi{1} & \mi{0} & \mi{0}
\cr \mi{0} & \mi{0} & \mi{0} & \mi{0} & \mi{0} & \mi{0} & \mi{0} &
\mi{1} & \mi{0} & \mi{0} \cr \mi{0} & \mi{0} & \mi{0} & \mi{0} &
\mi{0} & \mi{0} & \mi{0} & \mi{0} & \mi{1} & \mi{0} \cr \mi{0} &
\mi{0} & \mi{0} & \mi{0} & \mi{0} & \mi{0} & \mi{0} & \mi{0} & \mi{1}
& \mi{0} \cr \mi{0} & \mi{0} & \mi{0} & \mi{0} & \mi{0} & \mi{0} &
\mi{0} & \mi{0} & \mi{0} & \mi{1} \cr \mi{0} & \mi{0} & \mi{0} &
\mi{0} & \mi{0} & \mi{0} & \mi{0} & \mi{0} & \mi{0} & \mi{0} \cr },\
\ \L_2=\pmatrix{ \mi{0} & \mi{0} & \mi{1} & \mi{\hyp1} & \mi{0} &
\mi{0} & \mi{0} & \mi{0} & \mi{0} & \mi{0} \cr \mi{0} & \mi{0} &
\mi{0} & \mi{0} & \mi{0} & \mi{\hyp1} & \mi{0} & \mi{0} & \mi{0} &
\mi{0} \cr \mi{0} & \mi{0} & \mi{0} & \mi{0} & \mi{0} & \mi{0} &
\mi{1} & \mi{\hyp1} & \mi{0} & \mi{0} \cr \mi{0} & \mi{0} & \mi{0} &
\mi{0} & \mi{0} & \mi{0} & \mi{\hyp1} & \mi{0} & \mi{0} & \mi{0} \cr
\mi{0} & \mi{0} & \mi{0} & \mi{0} & \mi{0} & \mi{0} & \mi{0} & \mi{0}
& \mi{0} & \mi{0} \cr \mi{0} & \mi{0} & \mi{0} & \mi{0} & \mi{0} &
\mi{0} & \mi{0} & \mi{0} & \mi{\hyp1} & \mi{0} \cr \mi{0} & \mi{0} &
\mi{0} & \mi{0} & \mi{0} & \mi{0} & \mi{0} & \mi{0} & \mi{0} & \mi{1}
\cr \mi{0} & \mi{0} & \mi{0} & \mi{0} & \mi{0} & \mi{0} & \mi{0} &
\mi{0} & \mi{0} & \mi{\hyp1} \cr \mi{0} & \mi{0} & \mi{0} & \mi{0} &
\mi{0} & \mi{0} & \mi{0} & \mi{0} & \mi{0} & \mi{0} \cr \mi{0} &
\mi{0} & \mi{0} & \mi{0} & \mi{0} & \mi{0} & \mi{0} & \mi{0} & \mi{0}
& \mi{0} \cr }\ ,\eqn\matexamp
$$
where $\L_3=\shalf{\L_1}^3+\coeff32\L_1\L_2$, $\ \L_4=-\shalf{\L_1}^4
-\coeff52{\L_1}^2\L_2$ are not independent; the generators satisfy
the relations ${\L_1}^3\L_2 + \L_1{\L_2}^2 = {\L_2}^2 - {\L_1}^4 -
4{\L_1}^2\L_2=0$.

For general $k$, the ring generators satisfy certain relations that
are the analogs of the characteristic equation \chareq.
These relations are generated by solving the matrix system
$$
[x_1{\bf1}-\L_1]\cdot\Psi\ =\ [x_2{\bf1}-\L_2]\cdot\Psi\ = \ 0\ ,
\eqn\doublesys
$$
and integrate exactly to the superpotentials given in \genallW, up
to reparametrizations. More precisely, we find that the
superpotentials associated with the ring structure constants
$\L_{1,2}$ are given by $\wnk3k(x_1, \shalf x_2+\shalf{x_1}^2)$, and
we will consider (implicitly) only this parametrization of the
superpotentials in the following. In this parametrization,
the relevant recursion relation \relallW\ looks
$$
\wnk3k(x,0)\ =\ \coeff1{k+3}\big[\,\dx1 -
x_1\dx2\,\big]\wnk3{k+1}(x,0)\ .
\eqn\crucialrec
$$
Specifically, the first few such potentials for the models $\cph 2k$
are
$$
\eqalign{
\wnk30(x)\ &=\ \coeff16{x_1}^3+\shalf
x_1x_2\qquad\qquad\qquad\qquad({\rm trivial},\,c=0)\cr \wnk31(x)\
&=\ \coeff14{x_1}^4-\coeff14{x_2}^2+\shalf{x_1}^2x_2\qquad\qquad\
({\rm A_3},\,c=\coeff32)\cr \wnk32(x)\ &=\ \coeff15{x_1}^5-
x_1{x_2}^2\qquad\qquad\qquad\qquad\ ({\rm D_6},\,c=\coeff{12}5)\cr
\wnk33(x)\ &=\ \coeff13{x_2}^3-{x_1}^4x_2-2{x_1}^2{x_2}^2
\qquad\qquad\!({\rm J_{10}},\,c=3)\ ,\cr}
\eqn\expots
$$
where we also indicated the singularity types as well as the
central charges of the corresponding \nex2 theories. For $k=1$, the
\nex2 $W_3$ algebra truncates to the ordinary
\nex2 superconformal algebra, so that the $D_6$ model represents the
simplest non-trivial \nex2 $W_3$ minimal model with a non-null chiral
spin-3 current.

The components of the solution vectors to \doublesys\ are
given, up to reparametrization and vanishing relations, precisely by
the $W_3$-eigenpolynomials:
$$
\eqalign{
\{\Psi_i\}\ &=\ \big\{\,\Phi^{m_1,m_2}(x_1,\shalf
x_2+\shalf{x_1}^2),\,m_1+m_2\leq k\,\big\}\ =\ \rnk3k\ \cr &=\
\big\{\,1,x_1,\shalf({x_1}^2+x_2),
\shalf({x_1}^2-x_2),x_1x_2,{x_1}^3+x_1x_2,\dots \big\}\ .
\cr}\eqn\ringpolys
$$
This gives an explicit translation table between the weights of $\Xi$
and the LG polynomials, which are labelled by the
$\widehat{s\ell}(3)$
integrable highest weights $(m_1,m_2)$.

%%%%%%%%%%%%%%%%%%%%%%% %%%%%%%%%%%%%%%%%%%%%%%%%%%%%%%%
\chapter{Matrix formulation of integrable hierarchies}
\section{Dispersionless KdV hierarchy}
%%%%%%%%%%%%%%%%%%%%%%% %%%%%%%%%%%%%%%%%%%%%%%%%%%%%%%%

In order to present further below a matrix formulation of the
ordinary KdV hierarchy\foot{With ``KdV hierarchy'' we will always
mean the type $(1,k+2)$ generalized KdV hierarchy.}
\refs{\DS{,}\genDS{,}\krich}\ (pertaining to the LG models
$A_{k+1}\cong \cph 1k$), we will first review some facts about the
quasi-classical limit of the scalar Lax formulation.

\ni The scalar Lax operator is given by
$$
L(D,g) \ =\ \coeff1{k+2}D^{k+2} -
 \sum_{i=0}^{k} g_{i+2}(t)\,D^{k-i}\ ,
\eqn\sclax
$$
where $D\equiv\del_\xi$, and where $g_i$ are certain functions of the
KdV times $t_i$ and the cosmological constant $\xi$. (We will
restrict ourselves to the ``small phase space'', \ie, the
non-vanishing couplings correspond to the topological primary fields:
$t_i=0$ for $i\leq1$. A subscript will always denote the scaling
degree.) The functions $g(t,\xi)$ are determined by the flow
equations
$$
\del_{t_{i+2}}L(D,g)\ =\ \big[\,\O_{k+1-i},L\,\big](D,g)\ ,
\eqn\kdvflo
$$
$i=0,\dots,k$,  where
$$
\O_i(D,g)\ =\ \coeff1i\,\big((k+2)L\big)^{{i\over k+2}}_+(D,g)
\eqn\kdvham
$$
are the hamiltonians. The subscript ``$+$'' denotes, as usual,
the truncation to non-negative powers of $D$.
The associated linear system is \DS
$$
\big[\,\del_{t_{i+2}} - \O_{k+1-i}\,\big] \cdot \Psi_0\ =\ 0\ ,
\eqn\linsys
$$
together with
$$
L(D,g)\,\Psi_0\ =\ \coeff1{k+2}\,z\,\Psi_0\ ,
\eqn\scalspec
$$
where $\Psi_0$ is the Baker-Akhiezer function
and $z$ the spectral parameter.

\ni Boundary conditions are imposed via the string equation
$[L,D]=1$,
that is,
$$
D\,g_i \ = \d_{i,k+2}\ \ \ \longrightarrow\ \ \  \xi\equiv t_{k+2}\ .
\eqn\stringeq
$$
Since for the topological models $D$ never acts on $g_i\
(i\not=k+2)$, one may regard $D$ as a $c$-number variable, and call
it $x$. The Lax operator turns then into the LG superpotential of
type $A_{k+1}$:
$$
\wnk2k(x,g)\ =\ L(D\!\to\!x,g)\ .
\eqn\wlcorr
$$
(for convenience, we will write $W(x,g)\equiv(k+2)\,\wnk2k$ in the
following). This amounts to going to the dispersionless limit of the
KdV hierarchy \disp, where one replaces the conjugate variables
$(\xi,D)$ by $(\xi,x)$ and the commutator in \kdvflo\ by a Poisson
bracket, $\{\O,W\}\equiv \del_x\O\del_\xi W-\del_\xi \O\del_xW$. Note
that this limit is exact for the topological models \TFTmat, as long
as one restricts to the small phase space. Since $\del_\xi\O=0$ on
the small phase space hamiltonians, the flow equations become
$$
-\,\del_{t_{i+2}}W(x,g(t))\ =\ \del_x\,\O_{k+1-i}(x,g(t))\ .
\eqn\dispflo
$$
These equations determine the dependence of the couplings $g(t)$
on the KdV times $t_i$, which now have the interpretation as
flat coordinates on the LG deformation space. The flat fields,
\ie, the perturbed ring elements in a flat basis, are
\DVV
$$
\phi_i(x,t)\ =\ \del_x\,\O_{i+1}(x,g(t))\ ,\qquad\ \ \
i=0,\dots,k\ .
\eqn\kflat
$$
Note that one can define hamiltonians $\O_i$ for arbitrary integers
$i\geq1$; these describe flows associated with the gravitational
descendants \EYQ:
$$
\sigma_l(\phi_i)(x,t)\ =\
\del_x\,\O_{i+(k+2)l+1}(x,g(t))\ .
\eqn\gravflat
$$
The above operators are flat in the sense that the
associated Gauss-Manin connection
vanishes \refs{\flatcoa{,}\flatcob{,}\EYQ}.

\ni Note that one can write
$$
W(x,t)\ =\ \del_x V(x,t)\ ,
\eqn\prep
$$
so that the flow equations are equivalent to
$$
\O_{k+1-i}(x,t)\ =\ -\,\del_{t_{i+2}}V(x,t)\ .
\eqn\hamdefV
$$
Thus, one may regard the prepotential $V$ as a more fundamental
object, and indeed, $V$ is precisely the potential of the Kontsevich
model \Kons, which is the matrix model that underlies the topological
matter-gravity system.

%%%%%%%%%%%%%%%%%%%%%%% %%%%%%%%%%%%%%%%%%%%%%%%%%%%%%%%
\section{Matrix formulation}
%%%%%%%%%%%%%%%%%%%%%%% %%%%%%%%%%%%%%%%%%%%%%%%%%%%%%%%

In the spirit of the KP hierarchy, one may define \krich\ a
``pseudo-polynomial'', $K=: x+\sum c_l x^{-l}$, by
$$
K^{k+2}(x,x^{-1},g) \ = \ W(x,g)\ .
\eqn\kdef
$$
Accordingly, the LG field, viewed as a multi-valued functional, can
be
expanded in an infinite series:
$$
x(W(g))\ =\ K + \cO(K^{-1})\ .
\eqn\xKexp
$$
In terms of $K$, the hamiltonians \kdvham\ are given by
$$
\O_i(x,g)\ =\ \coeff1i\big(K^i(x,x^{-1},g)\big)_+ \ =:\ \coeff1i\,K^i
+
\sum_{l=1}^\infty b_l^i(g)\,K^{-l}\ ,
\eqn\khamdef
$$
where ``$+$'' denotes the truncation to non-negative powers of $x$ in
the expansion
$$
K^i(x,x^{-1},g)\ =:\ x^i + \sum_{l=2}^\infty c_l^i(g)\,x^{i-l}\ .
\eqn\kexp
$$
In terms of these quantities, the flow equations \dispflo\ are
equivalent to
$$
\del_{t_{i+2}}\,x(g(t))\ =\ \,\del_\xi\,\O_{k+1-i}(g(t))\ ,
\eqn\kflow
$$
which is actually part of the zero curvature system
$$
\big[\,\del_{t_{i+2}}-\O_{k+1-i},\, \del_{t_{j+2}}-\O_{k+1-j}\,\big]\
=\ 0\
\eqn\zerocur
$$
($\O_1\equiv x, \xi\equiv t_{k+2}$, $i,j=0,\dots,k$).
Note that the hamiltonians commute identically.

The above can given a matrix representation in the following way. One
first needs to find a $(k+2)\times (k+2)$ dimensional matrix Lax
operator\foot{We will use the term ``Lax operator''
for the derivative-free piece  $\cL_1$.}
such that the first order Drinfeld-Sokolov system
$$
\Big[\,D\bfone - \cL_1\,\Big]\cdot \Psi\ =\ 0
\eqn\firstorder
$$
reproduces, upon recursively solving for the components of $\Psi$,
the spectral equation \scalspec. It is well-known \DS\ that
this operator has the form
$$
\cL_1\ =\  \Lz_1 + Q_1(g)\ ,
\eqn\Ldef
$$
where $\Lz_1$ can be taken as
$$
\Lz_1\  =\ \pmatrix
{ 0 & 1 & 0 & \dots & 0 \cr
  0 & 0 & 1 & \dots & 0 \cr
  \vdots & \vdots & \vdots & \ddots &  \vdots \cr
0 & 0 & 0 & \dots & 1 \cr
z & 0 & 0 & \dots & 0 \cr}\ \ ,
\eqn\zstep
$$
and where $Q_1$ is usually taken to be a lower triangular matrix that
is determined only up to gauge transformations belonging to the
nilpotent subgroup $N^-$.

In the dispersionless limit, the spectral equation \scalspec\ becomes
$[W(x,g)-z]\,\Psi_0=0$, which is the equation that is supposed to be
obtained by recursively solving for the components of $\Psi$ in
$[x_1\bfone-\cL_1]\Psi=0$. Since this is precisely the characteristic
equation of $\cL_1$, which captures the gauge invariant content of
\firstorder, it follows that the matrix itself must satisfy
$$
W(\cL_1,g)\ =\ z\bfone\ .
\eqn\wident
$$
This ``superpotential spectral equation'' will prove crucial
for our purposes and can be taken as the definition of $\cL_1$ in
terms
of the \LG superpotential; it (non-uniquely) determines $Q_1(g)$.
The gauge freedom can be fixed by going to any particular gauge. The
choice that is most appropriate for us is however not given by taking
$Q_1$, as usual, to be a lower triangular matrix, but by taking $Q_1$
to belong \genDS\ to the Heisenberg subalgebra generated by $\Lz_1$
($Q_1$ is then lower triangular only up to $\cO(1/z)$). That is, we
have an infinite expansion
$$
\cL_1(g)\ =\ \Lz_1 + \sum_{l=1}^\infty q_l(g) (\Lz_1)^{-l}\ ,
\eqn\Lexpan
$$
whose coefficients can be computed from \wident\ in a recursive way.

Comparing with \xKexp, we see that $\cL_1$ is a matrix representation
of the LG field $x$ and that $\Lz_1$ is a matrix representation of
$K$. The defining equation \kdef\ of $K$ is trivially satisfied:
$$
K^{k+2}\ \equiv\ (\Lz_1)^{k+2}\ =\ z \bfone\ =\ W(\cL_1,g)\ .
\eqn\Ktriv
$$
We can thus interpret the foregoing equations as matrix equations,
and it is this form of the hierarchy that is most useful for our
generalization. In particular, the flow equations are
$$
\eqalign{ \del_{t_{i+2}}\,\O_{k+1-j}(g(t))\ &=\
\del_{t_{j+2}}\,\O_{k+1-i}(g(t))\ ,\cr \O_i(\cL_1(g),g)\ &\equiv\
\coeff1i(\Lz_1(\cL_1(g),\cL_1^{-1}(g),g))^i_+\ ,\qquad\
\O_1\equiv\cL_1\ .
}\eqn\matrixflows
$$
Obviously, the flows associated with
$\O_{l(k+2)}=\coeff1{l(k+2)}\,z^l\bfone$ are trivial, and correspond
to perturbations by the null operators $\sigma_l(\phi_{k+1})$.

%%%%%%%%%%%%%%%%%%%%%%% %%%%%%%%%%%%%%%%%%%%%%%%%%%%%%%%
\section{Heisenberg algebras and ``quantum'' chiral rings}
%%%%%%%%%%%%%%%%%%%%%%% %%%%%%%%%%%%%%%%%%%%%%%%%%%%%%%%

We notice that the matrix $\Lz_1$ \zstep\ figuring in the
$(1,k+2)$-type KdV hierarchy has the same form as the structure
constant matrix $\L_1$ \step\ of the $A_{k+1}$-type chiral rings
(apart from the spectral parameter $z$). In both cases, the matrix
represents a principally embedded $s\ell(2)$ step generator $I_+$.
Note, however, that despite of this similarity, the hamiltonians
$\O_i(t)=(\Lz_1)^i+\cO(t)$ that figure in the KdV equations are
$(k\!+\!2)\!\times\!(k\!+\!2)$-dimensional matrices, whereas the ring
structure constants $C_i(t)=(\L_1)^i+\cO(t)$ are only
$(k\!+\!1)\!\times\!(k\!+\!1)$ dimensional. The point is that the KdV
system does not only describe the matter sector, but also the null
state and gravitational sector of the theory (which is non-trivial
even for trivial matter where $k=0$). More precisely, the perturbed
version of \mateq\ is
$$
\Big[\,x{\bf1} - C_1(t)\,\Big] \cdot \Psi\ =\ 0\ ,
\eqn\mateqa
$$
and the solution vector components are the flat fields,
$\Psi_i \equiv \phi_i(x,t)$, $i=0,\dots,k$. One the other hand,
the KdV linear system \firstorder, \Lexpan\ in
the dispersionless limit is:
$$
\Big[\,x\bfone - \O_1(t,z)\,\Big]\cdot \Psi\ =\ 0\ .
\eqn\dfirstorder
$$
Here one obtains in a recursive manner the
flat fields $\phi_i$, $i=0,\dots,(k+1)$, plus all their
gravitational descendants, organized by $z$-series expansion:
$$
\Psi_i(x,t,z)\ =\ \sum_{l=0}^\infty z^{-l} \psi_{i,l}(x,t,z)
\eqn\Psiexp
$$
\vskip -.3 truecm \ni where
$$
\eqalign{
\psi_{i,l}(x,t,0)\ =\ -\del_z\psi_{i,l+1}(x,t,z) \ &=\
\coeff1{i+(k+2)l+1}\,\del_x\,\big(W(x,t)\big)^{{i+1\over k+2}+l}_+\cr
&\equiv\ \sigma_l(\phi_i)(x,t)\ .\cr}
\eqn\psigrav
$$
We thus see that the LG field $x$ can be represented either by the
ring structure constant $C_1$ or (formally) by the KdV Lax operator
$\cL_1\equiv\O_1$, depending on whether the space it acts upon is the
finite dimensional Hilbert space of the topological matter model, or
the infinite dimensional space of ``flat'' polynomials
$\sigma_l(\phi_i)$ of the matter-gravity system. To obtain the
Hilbert space of the matter-gravity system, one needs to mod out the
space of polynomials $\{\sigma_l(\phi_i)\}$ by the null fields,
$\sigma_l(\phi_{k+1}) \equiv(\del_x W)W^l$; it is therefore precisely
the null fields that account for the difference in the dimensions of
the matrices $C_1$ and $\O_1$.

Our plan is to make use of this correspondence between chiral
ring structure constants $C_i$ and hamiltonians $\O_i$ for the
generalization to $W$-gravity, by relating the hamiltonians
to certain {\it other}, higher dimensional chiral rings.

The relevant underlying algebraic structure of the hamiltonians,
$\O_i\equiv (\Lz_1)^i_+$, that gives the complete characterization
of the KdV flows (including the trivial ones), is given by
$$
\wth^+\ \equiv\ \big\{(\Lz_1)^m\,,\ m\in\ZZ_+\big\}\ .
\eqn\Hpospart
$$
This is the positive part of the infinite dimensional, maximally
commuting principal Heisenberg subalgebra $\cH\subset\widehat
{s\ell}(k\!+\!2)$ generated by \doubref\kac\genDS\
$$
\Lz_1\ \equiv\ \big(\!\!\sum_{{\rm{simple\atop
roots\ \alpha}} }\, E_\alpha \big) + z\, E_{-\psi}\ ,
\eqn\lzdef
$$
where $\psi$ denotes the highest root. Such an extension of a Lie
algebra to an affine algebra, via addition of $(-\psi)$ to the set of
simple roots, is well-known to play an important r\^ole in the
context of integrable perturbations. Indeed, $\Lz_1$ is identical to
the chiral ring structure constant $C_1(z)$ associated with the
following perturbed LG potential ``at one level higher'': \foot{This
perturbation is known to be quantum integrable and can be described
in terms of affine Toda theory \refs{\FMVW{,}\superpot{,}\ttstar.}}
$$
\wnk2{k+1}(x,t_{k+2}\!=\!z,t_l\!=\!0)\ =\
\coeff1{k+3}x^{k+3} - z\, x\ .
\eqn\wpert
$$
That is, {\it the underlying algebraic structure, $\cH^+$, of the
$A_{k+1}$ matter-gravity integrable system is that of a specifically
deformed chiral ring pertaining to the LG theory} $A_{k+2}$:
$$
\cH^+\ \cong\ \rnk2{k+1}(t_{k+2}\!=\!z,t_l\!=\!0)\ .
\eqn\HRrel
$$
Mathematically, the deformation by the spectral parameter $z$ is
precisely what deforms the cohomology ring \ringiso\ $H^*$ into the
quantum cohomology ring $QH^*$ \qucoho, whence
$$
\cH^+\ \cong\ QH^*_{\bar \del}\big({\rm CP}_{k+1},\IR\big)\ .
\eqn\QHrel
$$
(The word ``quantum'' indicates that the deformation of the classical
cohomology ring by the spectral parameter $z$ is precisely the effect
of the instanton corrections in a supersymmetric ${\rm CP}_{k+1}$
$\sigma$-model. From this viewpoint, the gravitational
descendant sectors correspond to the non-trivial instanton sectors of
the $\sigma$-model).

Most importantly, the spectral equation
\scalspec\ in the dispersionless limit is precisely the vanishing
relation of the LG theory \wpert\ at one level higher,
$$
\wnk2k(x,0)- \coeff1{k+2}z\ =\ \coeff1{k+2}\,\del_x\,
\wnk2{k+1}(x,z)\ ,
\eqn\wrel
$$
and this is what is at the heart of the matrix spectral equation
\wident, $W_{2,k}(\Lz_1)\equiv\coeff1{k+2}(\Lz_1)^{k+2}=\coeff1{k+2}
z
\bfone$.

We will take these facts, which are rather trivial here for one
LG variable, as starting points for our generalization.

%%%%%%%%%%%%%%%%%%%%%%% %%%%%%%%%%%%%%%%%%%%%%%%%%%%%%%%
\section{Generalization to $W_3$}
%%%%%%%%%%%%%%%%%%%%%%% %%%%%%%%%%%%%%%%%%%%%%%%%%%%%%%%

We are now in the position to formulate, tentatively, a
generalization of the dispersionless KdV hierarchy to several LG
variables. As we have seen in \hsubsect{2.2}, the chiral rings $\rnk
{n}k$ for $n\!>\!2$ are characterized by taking for $\Xi$ not the
defining representation, but the ($n\!-\!1$)th fundamental
representation of $SU(n+k-1)$. Thus, the most direct extension of the
KdV system would be to just consider matrix Lax operators in the
corresponding higher fundamental representations of $SU(n+k)$. As we
will see, this amounts to describe the integrable systems in terms of
perturbed chiral rings at one level higher\foot{Heuristically, one
may view this shift $k\to k+1$ as due to the well-known shift of the
$\widehat{s\ell}(n)$ integrable weights by the Weyl vector
$\rho_G$.}, $\rnk {n}{k+1}$, just like for $n\!=\!2$.

\ni Specifically, focusing on
$W_3$ at level $k$ associated with chiral rings $\rnk 3k$, we
consider the perturbed $SU(2)$ generator
$$
\Lz_1\
\equiv\ \L_1 + z\,\L_{-(k+2)}
\eqn\lzdefhi
$$
now in the $\shalf(k+2)(k+3)$ dimensional
${\lower2pt\copy111}\,$-representation of $SU(k+3)$. From
\hsubsect{2.2}\ we know that in this representation there exists an
additional, independent ring generator at grade two; requiring it to
commute with $\Lz_1$, we find
$$
\Lz_2\ =\ \L_2 + z\,\L_{-(k+1)}
%\ \not=\ (\Lz_1)^2
\ .\eqn\Lamtwo
$$
The generators $\Lz_1, \Lz_2$ are identical to the ring structure
constants corresponding to the following LG potential with
(integrable) perturbation:
$\wnk3{k+1}(x,z)\equiv\wnk2{k+1}(x,t_{k+3}\!=\!z,t_i\!=\!0)- \a\, z\,
x_1$ ($\a$ is some numerical factor that we will neglect in the
following). Using \crucialrec\ we can thus write:
$$
\wnk3k(x_1,x_2,0)-
\coeff1{k+3}z\ =\ \coeff1{k+3}\,[\,\del_{x_1}-x_1\del_{x_2}\,]
\,\wnk3{k+1}(x_1,x_2,z)\ .
\eqn\owrel
$$
This gives us the rationale for considering this construction, since
what \owrel\ means is that we have found matrices that satisfy a
generalized superpotential spectral equation
$$
\wnk3k(\Lz_1,\Lz_2,0)\ =\ \coeff1{k+3}\,z\,\bfone\ .
\eqn\crucial
$$
This is indeed preciseley what we have been looking for: $\Lz_1,
\Lz_2$ play the r\^ole of $K$ that represents the ($k\!+\!2$)-th root
of the ordinary, dispersionless Lax operator (cf., \Ktriv). In fact,
writing down such a matrix equation is non-trivial, since for large
$k$ the superpotential $\wnk3k(\Lz_1,\Lz_2,0)$ has an arbitrary
number of terms whose relative coefficients are fixed and correspond
to the correct, specific point in the LG moduli space. Note also that
by virtue of \relallW, equation \crucial\ can be immediately
generalized to all $W_n$.

\ni We now introduce LG couplings $g(t)$ according to
$$
\wnk3k(x,g)\ =\
\wnk3k(x_1,x_2,0)\,-
\!\!\!\sum_{m_1+m_2\leq k} g_{m_1,m_2}(t)\Phi_{m_1,m_2}(x_1,x_2)\ ,
\eqn\Wpert
$$
where $\Phi_{m_1,m_2}$ denotes the unperturbed ring elements
\ringpolys. Writing $W\!\equiv\!(k\!+\!3)\wnk3k$, the spectral
equation that generalizes \wident\ then looks
$$
W(\cL_1,\cL_2,g)\ =\ \,z\,\bfone\ ,
\eqn\newcrucial
$$
which involves perturbed Lax operators of the form
$$
\eqalign{
\cL_1(g,z)\ &=\ \Lz_1 + Q_1(g)\ \equiv\ \Lz_1 + \sum_{l,m}
q_{l,m}^1(g)\, (\Lz_1)^{-l}(\Lz_2)^{-m}\cr \cL_2(g,z)\ &=\ \Lz_2 +
Q_2(g)\ \equiv\ \Lz_2 + \sum_{l,m} q_{l,m}^2(g)\,
(\Lz_1)^{-l}(\Lz_2)^{-m}\ .\cr
}\eqn\Lexpanother
$$
The superpotential spectral equation \newcrucial\ corresponds to the
multi-variable analog of the condition
$W(\cL_1,g)_-\equiv(\Lz_1(\cL_1,{\cL_1}^{-1}))^{k+2}_-=0$, which
implements the reduction from the KP to the generalized KdV
hierarchy.

In order to obtain flow equations that will ultimately determine the
couplings $g(t)$ as functions of the flat coordinates, we need first
to construct appropriate hamiltonians. As we will see, the structure
of these hamiltonians is considerably more complicated as before
(c.f., \matrixflows).

The hamiltonians for a given integrable system are usually directly
given in terms of the underlying Heisenberg algebra: $\O=(\cH^+)_+$
\genDS. However, in the present context, where we consider
representations larger than the defining representation, there exist
in general more commuting matrices at a given grade than there are
elements of the Heisenberg algebra at this grade\foot{This
corresponds to ``type II hierarchies'' in the nomenclature of ref.\
\genDS.} (e.g., $(\Lz_1)^2$ does not belong to $\cH^+$). A priori,
any commuting matrix may give rise to a valid flow hamiltonian.
Therefore, the appropriate algebraic structure to look at is the {\it
enveloping} algebra $\wth^+$ of the principal Heisenberg algebra
$\cH^+$ that consists of the complete set of commuting matrices. It
is given by the perturbed chiral ring corresponding to the potential
$\wnk3{k+1}(x_1,x_2,z)$
\owrel
$$
\eqalign{
\wth^+\ &\equiv\ \big\{({\Lz_1})^{l_1}(\Lz_2)^{l_2}, \
l_i\geq0\,\big\}\ \cr &\cong\ \rnk3{k+1}(t_{k+3}\!=\!z,t_l\!=\!0)\ .
}\eqn\newHrel
$$
To characterize it mathematically, note that in precise analogy to
\QHrel, this ring is isomorphic to the ``quantum'' cohomology ring
\qucoho\ of the underlying grassmannian:
$$
\wth^+\ \cong\ QH^*_{\bar\del} \big(\,\coeff{SU(k+3)}{SU(k+1)\times
SU(2)\times U(1)}\,,\IR\,\big)\ .
\eqn\newQHrel
$$
The structure of $\wth^+$ \newHrel\ is more complicated than what the
notation
in \newHrel\ might suggest, in that $\Lz_1,\Lz_2$ satisfy non-trivial
relations. These relations, among which is the superpotential
spectral equation $W(\Lz_1,\Lz_2,0)=z\bfone$, truncate the Fock space
generated by $\Lz_1,\Lz_2$ such that there is only a finite number of
different Heisenberg algebra elements at any given grade.
Accordingly, we can write
$$
\eqalign{
\wth^+\ &=\ \bigoplus_{{\scriptstyle i=1,\dots,k+3}\atop{\scriptstyle
l\in\ZZ_+}}\wth_{i;l}\cr \wth_{i;l}\ &=\
\Big\{\,z^l\otimes\big[\oplus_{j=1}^{b_i}\Lz_{i,j}\,\big]\Big\}\cr
b_i\ &\equiv\ {\rm dim}\,\wth_{i;*}\ ,\ \ \ \sum b_i={\rm dim}\Xi\
,\cr
}\eqn\otherheidef
$$
where $\{\Lz_{i,j}\}$ denotes the set of Heisenberg algebra elements
at grade $i$. The fact that $b_i$ may be larger than one reflects the
possibility of having several degenerate flows at a given grade. (In
the following, we will drop the redundant label $l$ and take
$i\in\ZZ_+$.)

\ni Candidate commuting hamiltonians are given by
$$
\eqalign{
\O_{i,j}(\cL_1(g),\cL_2(g),g)\ =\
\big(&\Lz_{i,j}(\cL_1,{\cL_1}^{-1}\!\!,
\cL_2,{\cL_2}^{-1}\!\!,g)\big)_+\ ,\cr
&\Lz_{i,j}\in\wth^+\ ,
}\eqn\newhams
$$
where the subscript "+" indicates the truncation to positive powers
of both $\cL_1$ and $\cL_2$ in the Laurent expansion of $\Lz_{i,j}$.
The flow equations can then be written in the zero curvature form
$$
\eqalign{
\big[\,&D_{i,j}\,,D_{i',j'}\,\big]\ =\ 0\ ,\cr
&D_{i,j}\ \equiv\ {\del\over\del{t_{i,j}}} - \sum_{m=1}^{b_{k+4-i}}\,
Z_{i,j}^{(m)}\O_{k+4-i,m}(g(t)) \ , \cr
}\eqn\newfloeqs
$$
where $Z$ are normalization constants (we can always put
$Z_{1,j}^{(m)}=\delta_{j,1}\delta_{m,1})$. Whether these
equations really are meaningful or not is a question that
we cannot decide at this point. However, as we will explain in the
next section, the equations have indeed precisely the correct, unique
solutions $g(t)$ for the examples that we checked explicitly.

Note that the degeneracy of the hamiltonians is equal to or larger
than the degeneracy of the ring elements at the corresponding grade,
so that the equations are consistent; this can be deduced from
\newHrel. If the degeneracy of the hamiltonians $\O$ at a given grade
is larger than the degeneracy of the corresponding ring elements,
then there should be non-trivial relations between the coefficients
$Z$. This means that the normalization constants $Z$ should be
non-trivially determined by the flow equations (modulo trivial
rescalings of the $t_{i,j}$), and, as a consequence, that in general
not all Heisenberg algebra elements will generate independent
non-trivial flows. The situation is actually more involved than that,
as we will see momentarily.

The extra complication comes from the fact that the single equation
\newcrucial\ cannot fully determine all the coefficients $q(g)$ of
the matrices $\cL_{1,2}$ simultaneously; for this, one actually needs
to impose still another relation between $\cL_1$ and $\cL_2$. To
understand the correct choice of the extra relation, recall what the
r\^ole of the superpotential spectral equation \newcrucial\ is in
terms of the hamiltonian flows. The point is that the superpotential
represents a {\it constant} hamiltonian
$$
\big(\Lz_{k+3,1}\big)_+\ =\
W(\cL_1,\cL_2,g(t))_+\ =\ z\bfone\ =\ {\rm const}\ ,
\eqn\constflo
$$
which leads to a trivial flow. This trivial flow
is just the flow that is associated with the null field
$\del_{x_1}W$. It is then clear that the extra relation we need
to impose should correspond to a {\it constant}
hamiltonian that is related to the other null
field, $\del_{x_2}W$. This means that the correct extra relation
should have the form
$$
\hat W(\cL_1,\cL_2,\hat g(t))
 =\ {\rm const}\ \in\ \wth_{k+2;0}\ ,
\eqn\correctrel
$$
where $\hat W$ has degree $k+2$ and whose explicit form is a priori
undetermined and is supposed to be determined from the flow equations
\newfloeqs.

The other vanishing relation, $\del_{x_2}\wnk3{k+1}(\Lz)=0$,
that the matrices $\Lz_1, \Lz_2$ satisfy, implies that
$$
(\del_{x_2}\wnk3{k+1})(\cL_1(g),\cL_2(g))\attac{g=0}\ =\ 0\ ,
\eqn\nonconst
$$
and this means that the expansion of constant
Heisenberg algebra elements \newhams, which yields the hamiltonians,
is in general not unique, but rather is determined only up to terms
proportional to $\del_{x_2}\wnk3{k+1}$. For example, the following
hamiltonians are a priori defined only up to certain additional
free parameters $\a$:
$$
\eqalign{
\O_{k+2,j}(\cL_1,\cL_2,g,\hat g,\a)\ =\
&\big(\Lz_{k+2,j}(\cL_1,{\cL_1}^{-1}\!\!,
\cL_2,{\cL_2}^{-1}\!\!,g,\hat g)\big)_+\cr &+\
\a_{k+2,j}\,(\del_{x_2}\wnk3{k+1})(\cL_1,\cL_2)\ .\cr}
\eqn\freepara
$$
It turns out that, in general, some of the extra parameters $\a$ will
be fixed by the flow equations, and the dependence of the
hamiltonians on the surviving free parameters will just reflect the
freedom of adding terms proportional to vanishing relations to the
ring elements. That is, if $\phi(x,t)$ represents a perturbed chiral
ring element in a flat basis, then $\phi(x,t)+ \sum
p_a(x)(\del_{x_a}W(x,t))$ (for some polynomial $p$ of appropriate
degree) is a flat ring element as well.

The fact that the hamiltonians are not a priori completely defined in
terms of the Heisenberg algebra $\wth^+$, but initially depend on
some free parameters $Z,\a$, is in contrast to the ordinary KdV
system, where all hamiltonians \gravflat\ are completely determined
in terms of powers of $\Lz_1$ (up to normalization); this is
precisely because there is just one candidate hamiltonian at any
given grade.

As a consequence of these ambiguities, the hamiltonians \newhams\
associated with the gravitational descendants are not completely
determined in terms of the small phase space flow equations. Thus, in
order to construct the gravitational descendants, one needs to
consider the corresponding flow equations to fix these parameters.
Since
these equations involve couplings to the gravitational descendants,
one would need to step beyond the small phase space and the
dispersionless limit. This is beyond the scope of the present work,
and thus we cannot gain insight into the precise structure of the
$W$-gravitational descendants at this point.

%%%%%%%%%%%%%%%%%%%%%%% %%%%%%%%%%%%%%%%%%%%%%%%%%%%%%%%
\section{An example}
%%%%%%%%%%%%%%%%%%%%%%% %%%%%%%%%%%%%%%%%%%%%%%%%%%%%%%%

We like to demonstrate our ideas by
presenting as an example the model $\cph2k$ for $k=2$. We actually
also fully computed the model with $k=1$, which is however not that
much interesting, and obtained many terms of the Lax operators of the
model with $k=3$ as well. The latter model, though more interesting,
turns out to be technically too involved for us to be solved
completely, and we decided to refrain from writing down the correct,
but partial results we got. However, as far as we can see, the
treatment of this $k=3$ model is indeed completely parallel to $k=2$,
except for the appearance of a dimensionless modulus $t_0$, which is
at the origin of the technical complications.

The reader might object that since the model $\cph22$ is identical to
the twisted \nex2 minimal model of type $D_6$, the example will not
be very meaningful, and also that it has been already discussed in
the literature \refs{\DVV{,}\period{,}\taka}. However, this model
really is a true \nex2 $W_3$ minimal model with a non-zero spin-3
current, and exhibits non-trivial degenerate, extra flows precisely
according to the scheme that we have in mind. Moreover, our treatment
of this model is quite different from the treatment in the
literature, where the variable $x_2$ is eliminated at the expense of
introducing terms proportional to $(x_1)^{-1}$; this is reflected in
the appearance of inverse powers of $D$ in the differential scalar
Lax operator \refs{\DS,\BFFOW,\diFKu}.\foot {Such inverse powers in
scalar Lax operators derived from $[x_1-I_+-Q_1(g)]\Psi=0$ occur
whenever the representation $\Xi$ is reducible under the principal
$SU(2)$ subgroup $S$ \BFFOW. In our approach, negative powers do not
arise due to the additional equations $[x_p-\L_p-Q_p(g)]\Psi=0$,
$p>1$.} In our approach we do not eliminate $x_2$, and this is
precisely how our treatment of the model $\cph22$ captures the
essence of our construction, which is supposed to work for all $k$ in
an analogous way.

The Heisenberg algebra $\wth^+$ that is relevant for the model
$\cph22$ coupled to topological $W$-gravity (with potential
$\wnk32(x,z,0) = \coeff1{10}{x_1}^5-\shalf x_1{x_2}^2-\coeff45 z$) is
given by the perturbed chiral ring $\rnk 33(t_5=\coeff85z)$ of
$\cph23$, with potential $\wnk33(x,z,0) = \coeff13{x_2}^3 -
{x_1}^4x_2 - 2{x_1}^2{x_2}^2-\coeff85z\,x$. This ring is associated
with the ten dimensional representation of $SU(5)$ and is generated
by \goodbreak
$$\Lz_1=
\pmatrix{ \mi{0} & \mi{1} & \mi{0} & \mi{0} & \mi{0} & \mi{0} &
\mi{0} & \mi{0} & \mi{0} & \mi{0} \cr \mi{0} & \mi{0} & \mi{1} &
\mi{1} & \mi{0} & \mi{0} & \mi{0} & \mi{0} & \mi{0} & \mi{0} \cr
\mi{0} & \mi{0} & \mi{0} & \mi{0} & \mi{1} & \mi{0} & \mi{0} & \mi{0}
& \mi{0} & \mi{0} \cr \mi{0} & \mi{0} & \mi{0} & \mi{0} & \mi{1} &
\mi{1} & \mi{0} & \mi{0} & \mi{0} & \mi{0} \cr \mi{0} & \mi{0} &
\mi{0} & \mi{0} & \mi{0} & \mi{0} & \mi{1} & \mi{1} & \mi{0} & \mi{0}
\cr \mi{0} & \mi{0} & \mi{0} & \mi{0} & \mi{0} & \mi{0} & \mi{0} &
\mi{1} & \mi{0} & \mi{0} \cr \mi{0} & \mi{0} & \mi{0} & \mi{0} &
\mi{0} & \mi{0} & \mi{0} & \mi{0} & \mi{1} & \mi{0} \cr \mi{z} &
\mi{0} & \mi{0} & \mi{0} & \mi{0} & \mi{0} & \mi{0} & \mi{0} & \mi{1}
& \mi{0} \cr \mi{0} & \mi{z} & \mi{0} & \mi{0} & \mi{0} & \mi{0} &
\mi{0} & \mi{0} & \mi{0} & \mi{1} \cr \mi{0} & \mi{0} & \mi{0} &
\mi{z} & \mi{0} & \mi{0} & \mi{0} & \mi{0} & \mi{0} & \mi{0} \cr
 },\
\ \Lz_2=\pmatrix{ \mi{0} & \mi{0} & \mi{1} & \mi{\hyp1} & \mi{0} &
\mi{0} & \mi{0} & \mi{0} & \mi{0} & \mi{0} \cr \mi{0} & \mi{0} &
\mi{0} & \mi{0} & \mi{0} & \mi{\hyp1} & \mi{0} & \mi{0} & \mi{0} &
\mi{0} \cr \mi{0} & \mi{0} & \mi{0} & \mi{0} & \mi{0} & \mi{0} &
\mi{1} & \mi{\hyp1} & \mi{0} & \mi{0} \cr \mi{0} & \mi{0} & \mi{0} &
\mi{0} & \mi{0} & \mi{0} & \mi{\hyp1} & \mi{0} & \mi{0} & \mi{0} \cr
\mi{\hyp z} & \mi{0} & \mi{0} & \mi{0} & \mi{0} & \mi{0} & \mi{0} &
\mi{0} & \mi{0} & \mi{0} \cr \mi{z} & \mi{0} & \mi{0} & \mi{0} &
\mi{0} & \mi{0} & \mi{0} & \mi{0} & \mi{\hyp1} & \mi{0} \cr \mi{0} &
\mi{\hyp z} & \mi{0} & \mi{0} & \mi{0} & \mi{0} & \mi{0} & \mi{0} &
\mi{0} & \mi{1} \cr \mi{0} & \mi{0} & \mi{0} & \mi{0} & \mi{0} &
\mi{0} & \mi{0} & \mi{0} & \mi{0} & \mi{\hyp1} \cr \mi{0} & \mi{0} &
\mi{\hyp z} & \mi{0} & \mi{0} & \mi{0} & \mi{0} & \mi{0} & \mi{0} &
\mi{0} \cr \mi{0} & \mi{0} & \mi{0} & \mi{0} & \mi{\hyp z} & \mi{z} &
\mi{0} & \mi{0} & \mi{0} & \mi{0} \cr }\ ,
$$
We have $b_i=2$, and a basis for the enveloping algebra $\wth$ is
\goodbreak
$$
\eqalign{
\ \ \ \ \ \ \ \Lz_{1,1}\ &=\ \Lz_1\cr
\Lz_{1,2}\ &=\ \coeff1z(\Lz_1)^2(\Lz_2)^2\cr
\Lz_{2,1}\ &=\ \Lz_2\cr
\Lz_{2,2}\ &=\ (\Lz_1)^2\cr
\Lz_{3,1}\ &=\ \shalf[(\Lz_1)^3\!+\!3\Lz_1\Lz_2]\cr
\Lz_{3,2}\ &=\ \shalf[(\Lz_1)^3\,-\,\Lz_1\Lz_2]\cr
\Lz_{4,1}\ &=-\shalf[(\Lz_1)^2\Lz_2\!+\!(\Lz_2)^2]\cr
\Lz_{4,2}\ &=-\shalf[(\Lz_1)^2\Lz_2\!-\!(\Lz_2)^2]\cr
\Lz_{5,1}\ &=\ \coeff18[(\Lz_1)^5-5\Lz_1(\Lz_2)^2]\
%\equiv\ \coeff54 W(\Lz_1,\Lz_2)
\ =\ z\,\bfone\cr
\Lz_{5,2}\ &=\ \shalf[(\Lz_1)^3\Lz_2-3\Lz_1(\Lz_2)^2]\ ,\cr
}\eqn\Lamident
$$
which is complete up to arbitrary powers of the spectral parameter
$z$ (we use the convention that $\Lz_{i,1}\in\cH\subset\widehat
{s\ell}(5)$ for $i=1,\dots,4$). Accordingly, the Lax operators can be
expanded as
$$
\eqalign{
\cL_1(g,\hat g,z)\ &= \Lz_1 +
 \sum_{i=1}^5\sum_{j=1}^2\,q^1_{i,j}(g,\hat g,z)
\,\Lz_{i,j}\ \ \in\wth\cr
\cL_2(g,\hat g,z)\ &= \Lz_2 +
 \sum_{i=1}^5\sum_{j=1}^2\,q^2_{i,j}(g,\hat g,z)
\,\Lz_{i,j}\ \ \in\wth\ .\cr
}\eqn\Lexpans
$$
The coefficients $q$ depend on the coupling
constants $g,\hat g$ that figure in the superpotential
$$
W(x_1,x_2,g(t))\ =\ \coeff1{10}{x_1}^5-\shalf x_1{x_2}^2 -
g_1(t){x_1}^4 - \dots - g_*(t) x_2 - g_5(t) \ ,
\eqn\Wpert
$$
and in the extra relation
$$
\hat W(x_1,x_2,\hat g)\ =\ \hat g_0^{(1)}{x_1}^4 + \hat g_0^{(2)}
{x_1}^2x_2 + \hat g_0^{(3)} {x_2}^2+ \hat g_1(t){x_1}^3 - \dots -
\hat g_4(t) \
\eqn\exr
$$
(we choose here as ring basis not the one given in \ringpolys, but
the basis used in \refs{\DVV{,}\period{,}\taka}). It turns out that
writing the Lax operators and other hamiltonians in terms of a large
number of unknowns takes an enormous amount of space. Therefore, we
will write all these quantities directly in terms of the solutions
$g(t)$. We will in addition suppress the dependence on the parameter
$t_1$, which is not important for demonstrating the existence of
consistent, non-trivial degenerate flows. We emphasize, however, that
we did make the most general ansatz and indeed found flat coordinates
as {\it unique}\foot{Unique up to the freedom of adding trivial
vanishing relations to the flat ring elements.} solution of the flow
equations.

\ni Written in terms of the flat coordinates, the superpotential
spectral equation is
$$
\eqalign{
W(\cL_1,\cL_2,t)\ &=\ \coeff1{10}{\cL_1}^5 - \shalf
{\cL_1}{\cL_2}^2\cr
&- t_{2} {{\cL_1}^3} - t_{3} {{\cL_1}^2} + \coeff{1}{2} \left( 3
{{t_{2}}^2} - 2 t_{4} \right) {\cL_1} - t_{*} {\cL_2} +  t_{2}
t_{3} - t_{5} \cr &=\ \coeff45 \,z\,\bfone\ .
}\eqn\firstrel
$$
The extra relation, as determined by the flow equations, turns out to
be
$$
\eqalign{
\hat W(\cL_1,\cL_2,t)\ &=\ \coeff{3}{16} {{\cL_1}^4} -\coeff{3}{8}
{{\cL_1}^2} {\cL_2} - \coeff{1}{16} {{\cL_2}^2}\cr &-\coeff{1}{8}t_2
{{\cL_1}^2} + \coeff{1}{8}t_2 {\cL_2} - t_{*} {\cL_1}+\coeff{3
}{16}{{t_{2}}^2} - \coeff{t_{4}}{2} \cr &=\ \Lz_{4,1} +
\coeff54\Lz_{4,2}\
.\cr
}\eqn\firstrel
$$
These equations imply for the coefficients $q(t,z)$ in \Lexpans:
\def\orderz#1{ + {\cal O}(\!z^{-#1}\!)}
$$
\eqalign{
q^1_{1,1}\ &=\ \coeff{1}{2z} \left( 4 t_{2} t_{3} - 2 t_{5} + 13
t_{2} t_{*} \right) \orderz2\cr q^1_{1,2}\ &=-\coeff{1}{4z} \left( 9
t_{2} t_{3} - 7 t_{5} + 43 t_{2} t_{*} \right)\orderz2\cr q^1_{2,1}\
&=\ \coeff{1}{8z} \left( 26 {{t_{2}}^2} + 9 t_{4} \right)\orderz2\cr
q^1_{2,2}\ &=\ \coeff{1}{8z} \left( 8 {{t_{2}}^2} + 3 t_{4}
\right)\orderz2\cr q^1_{3,1}\ &=-\coeff{1}{4z} \left( 2 t_{3} - t_{*}
\right)\orderz2\cr q^1_{3,2}\ &=\ \coeff{1}{4z} t_{3} \orderz2\cr
q^1_{4,1}\ &=\ \coeff1{z} t_{2} - \coeff{1}{16z^2} \left( 709
{{t_{2}}^2} t_{3} + 27 t_{3} t_{4} - 112 t_{2} t_{5} + 1011
{{t_{2}}^2} t_{*} + 99 t_{4} t_{*} \right) \orderz3\cr q^1_{4,2}\ &=\
\coeff{1}{16z^2} \left( 486 {{t_{2}}^2} t_{3} + 21 t_{3} t_{4} - 64
t_{2} t_{5} + 612 {{t_{2}}^2} t_{*} + 59 t_{4} t_{*} \right)
\orderz3\cr q^1_{5,1}\ &=-\coeff{1}{64z^2} \left( 2360 {{t_{2}}^3} +
21 {{t_{3}}^2} + 488 t_{2} t_{4} - 278 t_{3} t_{*} + 17 {{t_{*}}^2}
\right)\orderz3\cr q^1_{5,2}\ &=-\coeff{1}{64z^2} \left( 1096
{{t_{2}}^3} + 19 {{t_{3}}^2} + 232 t_{2} t_{4} - 130 t_{3} t_{*} + 7
{{t_{*}}^2} \right) \orderz3\cr
}\eqn\qcoeffone
$$
and
$$
\eqalign{
q^2_{1,1}\ &=\ \coeff{1}{4z} \left( 240 {{t_{2}}^3} + 3 {{t_{3}}^2} +
56 t_{2} t_{4} - 23 t_{3} t_{*} + {{t_{*}}^2} \right)\orderz2\cr
q^2_{1,2}\ &=-\coeff{1}{16z} \left( 1480 {{t_{2}}^3} + 3 {{t_{3}}^2}
+ 336 t_{2} t_{4} - 142 t_{3} t_{*} + 7 {{t_{*}}^2}
\right)\orderz2\cr q^2_{2,1}\ &=\ -\coeff{1}{4z} \left( 20 t_{2}
t_{3} - 9 t_{5} + 48 t_{2} t_{*} \right)\orderz2\cr q^2_{2,2}\ &=\
\coeff{3}{4z} \left( t_{5} - 4 t_{2} t_{*} \right)\orderz2\cr
q^2_{3,1}\ &=- \coeff{1}{4z} \left( 25 {{t_{2}}^2} + 4 t_{4}
\right)\orderz2\cr q^2_{3,2}\ &=\ \coeff{1}{4z} \left( 9 {{t_{2}}^2}
+ 2 t_{4} \right)\orderz2\cr q^2_{4,1}\ &=-\coeff1{z} t_{*}
\orderz2\cr q^2_{4,2}\ &=\ \coeff{1}{2z} \left( 3 t_{3} - t_{*}
\right)\orderz2\cr q^2_{5,1}\ &=\ \coeff{1}{16z^2} \left( 1469
{{t_{2}}^2} t_{3} + 87 t_{3} t_{4} - 244 t_{2} t_{5} + 2019
{{t_{2}}^2} t_{*} + 151 t_{4} t_{*} \right)\orderz3\cr q^2_{5,2}\ &=
-\coeff1{z} t_{2} + \coeff{1}{16z^2} \left( 589 {{t_{2}}^2} t_{3} +
33 t_{3} t_{4} - 116 t_{2} t_{5} + 919 {{t_{2}}^2} t_{*} + 69 t_{4}
t_{*} \right) \orderz3\cr
}\eqn\qcoefftwo
$$
The flow equations \newfloeqs\ can be
written as ($I\equiv (2,\dots,5,*)$, $\O_{6-*}\equiv \O_*$):
$$
\del_{t_I}\O_{6-J}(\cL_1(t),\cL_2(t),t)\ =\
\del_{t_J}\O_{6-I}(\cL_1(t),\cL_2(t),t)\ ,
\eqn\dsixfloeqs
$$
and are satisfied by the matrix-valued hamiltonians $\O_I \equiv
\O_I(\cL_1(t), \cL_2(t), t)$:
$$
\eqalign{
\O_1\ &=\ \big(\Lz_{1,1}\big)_+\ =\ \cL_1 \cr \O_2\ &=\ \big(\shalf
\Lz_{2,1}\big)_+\ =\ \shalf\cL_2 \cr \O_3\ &=\ \big(\coeff16
\Lz_{3,1}+\shalf \Lz_{3,2}\big)_+\ =\ \coeff13 {\cL_1}^3 - {\cL_1}
t_2 -\coeff58t_3-\coeff18t_*\cr \O_*\ &=\ \big(\coeff16
\Lz_{3,1}-\coeff32 \Lz_{3,2}\big)_+\ =\ {\cL_1} {\cL_2} -\coeff{2}{3}
{{\cL_1}^3} + \coeff{15 }{8}t_{*} -\coeff{1}{8}t_{3}\cr \O_4\ &=\
\big(\coeff12 \Lz_{4,1}+\coeff34 \Lz_{4,2}\big)_++\a_1\big(\Lz_{4,1}
+ \coeff54\Lz_{4,2}\big)\ \cr &=\ \coeff{3}{16} {{\cL_1}^4}+
\coeff{1}{8} {{\cL_1}^2} {\cL_2} - \coeff{1}{16} {{\cL_2}^2}
+\coeff{31 }{16}{{t_{2}}^2} -\coeff{9 }{8}t_{2} {{\cL_1}^2} -\coeff{3
}{8}t_{2} {\cL_2} - t_{3} {\cL_1} \cr &\ \ + \a_1 \hat
W(\cL_1,\cL_2,t)\ .
}\eqn\hams
$$
Note that $\O_3, \O_*$ represent two degenerate flows, as advertised,
and that $\a_1$ is a free parameter which corresponds to the freedom
of adding a constant matrix to $\O_4$. \goodbreak

%%%%%%%%%%%%%%%%%%%%%%% %%%%%%%%%%%%%%%%%%%%%%%%%%%%%%%%
\section{Prepotential and scalar Lax system}
%%%%%%%%%%%%%%%%%%%%%%% %%%%%%%%%%%%%%%%%%%%%%%%%%%%%%%%

So far we considered the flow equations in matrix form,
cf., \matrixflows, \newfloeqs. There is also a scalar version of
these equations, which is obtained by replacing $\cL_i$ by
the LG fields $x_i$. Because of the dependence
of $\cL_i$ on the coordinates $t$, the form of the
scalar equations is different from their matrix versions.

We find for the examples for $W_3$-gravity that we computed
explicitly a structure that is very similar to the one of ordinary
gravity. For instance, for the $k=2$ model discussed in the previous
section (and similarly for the other examples), we find that the
scalar
Lax system can be written as:
$$
\eqalign{
-\del_{t_I}W(x_1,x_2,t)\ &=\
 \big\{\,\O_{6-I}(x_1,x_2,t),\,W\,\big\}\cr
-\del_{t_I}\hat W(x_1,x_2,t)\ &=\
 \big\{\,\O_{6-I}(x_1,x_2,t),\,\hat W\,\big\}\ ,\cr
}\eqn\poissflo
$$
with the Poisson bracket
$$
\eqalign{
\big\{\,\O\,,L\,\big\}\ \equiv\ \
&[(\del_{t_5}\O)(\del_{x_1} L) - (\del_{x_1}\O)(\del_{t_5} L)]\cr +
2\,
&[(\del_{t_4}\O)(\del_{x_2} L) - (\del_{x_2}\O)(\del_{t_4} L)]\ .\cr
}\eqn\Poissdef
$$
This corresponds to two pairs of conjugate variables, $(t_5,x_1)$ and
$(t_4,\shalf x_2)$. The equations \poissflo\ can be viewed as the
dispersionless limit of a differential scalar system of the form
$$
\eqalign{
-\del_{t_I}L(D_1,D_2,t)\ &=\
 \big[\,\O_{6-I}(D_1,D_2,t),\,L\,\big]\cr
-\del_{t_I}\hat L(D_1,D_2,t)\ &=\
 \big[\,\O_{6-I}(D_1,D_2,t),\,\hat L\,\big]\ ,\cr
}\eqn\poissfloL
$$
if one imposes as string equations
$$
\eqalign{
D_1\,t_5\ &=\ 1 \cr
D_2\,t_4\ &=\ 2 \ .\cr
}\eqn\newstringeq
$$
Presumably, the system \poissfloL\ gives the complete
characterization
of the $W_3$-matter-gravity model beyond the small phase space.

The terms in \poissflo\ that are proportional to $\del_{x_i} W$
precisely cancel out the dependence on the free parameter $\a_1$
in the hamiltonian $\O_4$ \hams. If one puts this parameter to
zero, one can write the scalar equations in a simpler form:
$$
\eqalign{
-\del_{t_I}W(x_1,x_2,t)\ &=\ (\del_{x_1} + 2 x_1
\del_{x_2})\,\O_{6-I}(x_1,x_2,t)\cr
-\del_{t_I}\hat W(x_1,x_2,t)\ &=\
 \del_{x_2}\,\O_{6-I}(x_1,x_2,t)\cr
}\eqn\scalflo
$$
which is the two-variable generalization of \dispflo. The first
equation gives the flat fields $\phi_{5-I}\equiv-\del_{t_{I}}W$
directly in terms of the hamiltonians.

Note that the equations \scalflo\ are precisely the
integrability conditions for the existence of a prepotential $V$ with
$$
\eqalign{
W(x,t)\ &=\ (\del_{x_1} + 2 x_1 \del_{x_2})\,V(x,t)\cr
\hat W(x,t)\ &=\ \del_{x_2}V(x,t)\cr
\O_I(x,t)\ &= -\,\del_{t_{6-I}}V(x,t) \ ,\cr
}\eqn\Vdef
$$
which is given by:
$$
\eqalign{
V(x,t)\ &=\ \coeff{3}{16} {{x_1}^4} {x_2}-\coeff{11}{240}
{{x_1}^6} - \coeff{3}{16} {{x_1}^2} {{x_2}^2} - \coeff{1}{48}
{{x_2}^3} - \coeff{3 }{16} t_{2}{{x_1}^4} - \coeff{1}{8}t_{2}
{{x_1}^2} {x_2} \cr &+ \coeff{9 }{16}{{t_{2}}^2} {{x_1}^2} +
\coeff{1}{16} \left( 3 {{t_{2}}^2} - 8 t_{4} \right) {x_2} + \left(
t_{2} t_{3} - t_{5} \right) {x_1} - \coeff{1}{3} \left( t_{3} - 2
t_{*} \right) {{x_1}^3} \cr &- t_{*} {x_1} {x_2}+ \coeff{1}{16}t_{2}
{{x_2}^2} + \coeff{5 }{16}{{t_{3}}^2}-\coeff{31 }{48}{{t_{2}}^3} +
\coeff{1}{8}t_{3} t_{*} - \coeff{15 }{16}{{t_{*}}^2}\ .
}
\eqn\Vexp
$$
The fact that
$$
\eqalign{
-\del_{t_5}V\ &=\ x_1 \cr
-\del_{t_4}V\ &=\ \shalf x_2\cr
}\eqn\newstringeq
$$
just expresses that $(t_5,x_1)$ and $(t_4,\shalf x_2)$ are
conjugate pairs. Moreover, note that the scalar spectral equations
can be thought of as variations of the prepotential:
$$
\delta_r\,V(x,t)\ =\ 0\ ,
\eqn\variations
$$
where
$$
\eqalign{
\delta_1\ &=\ \del_{x_1} + 2 x_1 \del_{x_2}\cr
\delta_2\ &=\ \del_{x_2}\cr
}\eqn\killing
$$
are nothing but Killing vectors of the ``$W_3$-plane'' \OSSPvN.
This hints at an interesting underlying geometrical structure.

It is fascinating to speculate that, in analogy to ordinary gravity
\prep, $V$ might be the potential of an appropriate underlying,
generalized two-matrix Kontsevich \Kons\ (or Kazakov-Migdal \kamig)
model.

%%%%%%%%%%%%%%%%%%%%%%% %%%%%%%%%%%%%%%%%%%%%%%%%%%%%%%%
\chapter{Discussion and Outlook}
%%%%%%%%%%%%%%%%%%%%%%% %%%%%%%%%%%%%%%%%%%%%%%%%%%%%%%%

We proposed in this paper a construction of integrable systems that
extends the dispersionless, $(1,k+2)$ type generalized KdV hierarchy
to several variables $x_i$. We focused mainly on two variables, which
corresponds to topological $W_3$-gravity coupled to matter, but it is
pretty obvious how to generalize this to $M$ variables.

A key point was to represent the superpotential spectral equations
as matrix relations:
$$
W(\cL_1, \cL_2,\dots,g)=z\,\bfone\
\eqn\supspeceqs
$$
(these equations are the multi-variable analogs of the constraint
that implements the reduction from the KP to the KdV
hierarchy). That these matrix relations happen to be identical to
chiral ring vanishing relations of certain {\it other} LG theories
was very
convenient and allowed us to realize them in terms of known chiral
ring structure constants. Accordingly, the hamiltonians can be
constructed in terms of the chiral rings of these other LG theories,
which are ``at one level higher''. In practice, one considers
Drinfeld-Sokolov matrix systems associated with the $(n\!-\!1)$-th
fundamental representation of $SU(n\!+\!k)$ (for models
$\cM^{(n)}_{1,n+k}$). In such representations, there exist in general
more commuting matrices at a given grade than there are powers of the
$s\ell(2)$ step generator, $\L_1$, and this leads to the
possibility of having more commuting flows as compared to the
ordinary generalized KdV hierarchy.

In mathematical terms, the underlying algebraic structure of the
relevant Heisenberg algebras is that of ``quantum'' cohomology rings
of grassmannians, $\wth^+\cong QH^*_{\bar\del}({SU(n+k)\over
SU(n-1)\times SU(k+1)\times U(1)},\IR)$, and the superpotential
spectral equations \supspeceqs\ are specific relations in these
rings.

So far, $W$-algebras were most often discussed \wconstr\ in the
context of ordinary gravity coupled to matter and not in the context
of higher $W_n$-gravities. For ordinary topological gravity coupled
to matter, which is described by the model $\cph1k$, the basic
underlying structure is the one of \lfig\figone, which is essentially
the positive affine weight space of $\hat{s\ell}(2)$. The picture may
simultaneously represent the chiral ring $\rnkg2k$ of the topological
LG model, the spectrum of KdV hamiltonians $L^{{(i+1)/(k+2)}}_+$, the
$W_{k+2}$-constraints of the matrix model, and also the classical
$W_{k+2}$ Poisson bracket algebra of the KdV hierarchy (note that
these classical $W$-algebras have nothing to do with $W$-gravity.)
The
important point is that all these objects are not truly independent,
but live in the enveloping algebra of a {\it single generator} (which
is,
morally speaking, given by an $s\ell(2)$ step generator).

The generalization to $W_n$-gravity coupled to matter discussed in
the present paper proceeds in an orthogonal direction and is
essentially a generalization to $M$ {\it independent generators},
where
$M\!\equiv\!n\!-\!1$ counts the number of independent generators of
the $W_n$ algebra. The situation for $n\!=\!3$ can be schematically
depicted as in \lfig\figtwo. That is, $M$ also counts the number of
independent LG fields (which generate the chiral rings $\rnk nk$),
the
number of gravitational descendant generators $\ga i$, the number of
independent \nex2 $W_n$ supercurrents that generate the chiral
algebra of the coset models $\cph Mk$, the number of Heisenberg
algebra generators $\Lz_i$ and the number of independent derivatives
in the differential scalar Lax operators $L(D_i)$ (cf., \poissfloL).
Moreover, we conjecture that there exist generalized Kontsevich
\Kons, or Kazakov-Migdal \kamig\ type matrix models that give an
equivalent description of the $W_n$-matter-gravity theories, where
$M$ is the number of matrices. These models are supposedly defined in
terms of prepotentials $V$ like \Vexp. If this
were true, one would have a matrix model description for at least
some models that are directly relevant for string compactification
(in particular, the models $\cph4{15},\cph59$ and $\cph67$
have $c\!=\!9$).

What these $M$ generators have in common is that they are
graded like the exponents $(1,2,3,...,M)$ (sometimes shifted by one
or two units), and this is ultimately inherited from the principal
embedding $s\ell(2)\hookrightarrow s\ell(n)$. We believe that our
findings can be generalized to other embeddings of $s\ell(2)$ into
Lie algebras, which would give rise more general kinds of gradations.

A generalization of Drinfeld-Sokolov systems with different than
principal gradations was considered before by refs.\
\doubref\genDS\othergrad, and one might ask about the relationship of
this to our work. Indeed one may view the generators $\L_i$, when
taken in some higher fundamental representation, also as generators
in the fundamental representation of some larger group with
appropriately chosen gradation $s$ (where $I_0=\sum s_i\lambda_i\cdot
H$). For example, the matrices in \matexamp, which represent $\L_1$
and $\L_2$ in the 10 dimensional representation of $SU(5)$ with
principal gradation $s=(1,1,1,1)$, can also be interpreted as
matrices in the fundamental representation of $SU(10)$ with gradation
$s=(1,1,0,1,0,1,0,1,1)$. However, the total grade,
$$
d_s\ =\ N_s (z\del_z) + \sum s_i\lambda_i\cdot H\ ,
$$
involves in addition the grade of the spectral parameter $z$, which
is given by $N_s=1+\sum s_i$ (for $SU(m)$). Obviously, the grade of
the spectral parameter in our construction (in the example, $[z]=5$)
is different from the corresponding grade of \genDS\ ($[z]=7$). This
means that the matrices $\Lz_i$ cannot be the same in the two
approaches (although they agree for $z=0$), and that the Heisenberg
algebras differ. It would be interesting to see how the work of
\genDS\ can be generalized such as to include the construction
discussed in the present paper.

We approached the problem of integrable systems pertaining to
$W$-gravity from a very specific viewpoint, namely by relating
dispersionless Lax operators with certain chiral rings. It is quite
clear that we just barely scratched the surface, and many important
questions are left unanswered. Among these open problems is a proof
that our method works in general and indeed describes topological
$W$-gravity including gravitational descendants, and this would
probably require to first develop an appropriate kind of intersection
theory of the correlation functions.

Furthermore, one would also like to go beyond the dispersionless
limit, in order to describe more general, non-topological models of
type $(p,q)$. As indicated in \hsubsect{3.6}, this would
amount to introducing an independent, commuting derivative for each
LG field, such that the Kazama-Suzuki superpotential turns into a
differential Lax operator: $W(x_1,x_2,\dots,x_M,g) \to
L(D_1,D_2,\dots,D_M,g)$. This scalar differential operator would be
associated to a matrix system of the form
$$
\eqalign{
\Big[\,D_1\bfone - \cL_1(g)\,\Big]\cdot \Psi\ &=\ 0\cr
\Big[\,D_2\bfone - \cL_2(g)\,\Big]\cdot \Psi\ &=\ 0\cr
 \vdots\qquad\qquad&\cr
\Big[D_{\!M}\bfone\! - \!\cL_{\!M}(g)\Big]\cdot \Psi\ &=\ 0\ ,\cr
}\eqn\newfirstorder
$$
with $\cL_p(g)=\Lz_p +Q_p(g)$, in generalization of \mateq\ and
\firstorder. The appearance of several kinds of derivatives would be
a
direct manifestation of $W$-geometry, via association with
coordinates in a ``$W$-superspace'', \ie, $D_1\to \del_{z_1}=L_{-1},
D_2\to \del_{z_2}=(W_3)_{-2}$, and so on.

We believe that the ideas presented in this paper can be
applied to a variety of related problems, presumably to all problems
that are based on the principal subgroup $SU(2)$, signalled by the
appearance of the familiar matrix \step. For example, we expect that
the construction of singular vectors of the Virasoro algebra given in
\virnul\ can be adapted to $W$-algebras by making use of equations
similar to \newfirstorder.

\vskip -1cm
%%%%%%%%%%%%%%%%%%%%%%% %%%%%%%%%%%%%%%%%%%%%%%%%%%%%%%%
\ack
%%%%%%%%%%%%%%%%%%%%%%% %%%%%%%%%%%%%%%%%%%%%%%%%%%%%%%%

I thank for discussions: Ioannis Bakas, Tim Hollowood, Elias
Kiritsis, Jan Louis, Andrej Marshakov, Luis Miramontes, Greg Moore,
Dennis Nemeschansky, Alexei Semikhatov, Alexander Sevrin, Erik
Verlinde, Gerard Watts, Jean-Bernard Zuber, and Nick Warner in
particular for ref.\ \KosB. In addition, I would like to thank the
Institute for Theoretical Physics in Santa Barbara for a very
pleasant stay, during which part of the present work was done. This
part of research was supported by the National Science Foundation
under Grant No.\ PHY89-04035.

\goodbreak%\vfil\eject

\refout
\end